\documentclass[12pt,tightenlines,preprint,eqsecnum,floats,aps,amsmath,amssymb,nofootinbib,prd,superscriptaddress,showpacs]{revtex4}

\usepackage{color}
\usepackage{graphicx}

\usepackage{amsfonts}
\def\be{\nopagebreak[3]\begin{equation}}
\def\ee{\end{equation}}
\def\ba{\nopagebreak[3]\begin{eqnarray}}
\def\ea{\end{eqnarray}}
\def\d{{\rm d}}
\def\b{{\tt b}}
\def\B{{}_{\rm B}}
\def\la{\lambda}
\def\d{{\rm d}}

\def\ihalf{\f{i}{2}}

\def\co{\sqrt{12 \pi G}}

\def\WDW{WDW\,\,}

\def\H{{\cal H}}

\def\Hp{\H_{\rm phy}}

\def\Hpwdw{\H_{\rm phy}^{\rm wdw}}

\def\V{{\cal V}}

\def\lp{{\ell}_{\rm Pl}}

\def\e{\mathring{e}}
\def\o{\mathring{\omega}}
\def\q{\mathring{q}}

\newcommand{\ket}[1]{\ensuremath{|#1\rangle}}

\newcommand{\rcr}{\rho_{\mathrm{crit}}}
\newcommand{\rsup}{\rho_{\mathrm{sup}}}

\newcommand{\p}{\partial}
\newcommand{\pf}{p_{(\phi)}}

\newcommand{\f}{\frac}

\def\phy{\mathrm{phy}}
\def\no{\nonumber}

\def\pphi{p_{(\phi)}}
\def\pphix{\langle \hat{p}_{(\phi)} \rangle}
\def\vex{\langle \hat V|_\phi \rangle}

\def\f{\frac}

\def\ul{\underline}
\def\ut{\tilde}
\def\t{\tilde}

\begin{document}
\preprint{\vbox{\baselineskip=12pt \rightline{IGC-07/10-01}
\rightline{PI-QG-61} }}
\title{Robustness of key features of loop quantum cosmology}
\author{Abhay Ashtekar}
\email{ashtekar@gravity.psu.edu} \affiliation{Center for Fundamental
Theory, Institute for Gravitation and the Cosmos, The Pennsylvania
State University, University Park PA 16802, USA}
\author{Alejandro Corichi}\email{corichi@matmor.unam.mx}
\affiliation{Instituto de Matem\'aticas, Unidad Morelia, Universidad
Nacional Aut\'onoma de M\'exico, UNAM-Campus Morelia, A. Postal
61-3, Morelia, Michoac\'an 58090, Mexico} \affiliation{Center for
Fundamental Theory, Institute for Gravitation and the Cosmos, The
Pennsylvania State University, University Park PA 16802, USA}
\author{Parampreet Singh}
\email{psingh@perimeterinstitute.ca}
\affiliation{Perimeter Institute for Theoretical Physics,
31 Caroline Street North, Waterloo, Ontario N2L 2Y5, Canada}
\affiliation{Center for Fundamental
Theory, Institute for Gravitation and the Cosmos, The Pennsylvania
State University, University Park PA 16802, USA}

\begin{abstract}

Loop quantum cosmology of the $k$=0 FRW model (with a massless
scalar field) is shown to be exactly soluble if the scalar field is
used as the internal time already in the classical Hamiltonian
theory. Analytical methods are then used i) to show that the quantum
bounce is generic; ii) to establish that the matter density has an
absolute upper bound which, furthermore, equals the critical density
that first emerged in numerical simulations and effective equations;
iii) to bring out the precise sense in which the Wheeler DeWitt
theory approximates loop quantum cosmology and the sense in which
this approximation fails; and iv) to show that discreteness
underlying LQC is fundamental. Finally, the model is compared to
analogous discussions in the literature and it is pointed out that
some of their expectations do not survive a more careful
examination. An effort has been made to make the underlying
structure transparent also to those who are not familiar with
details of loop quantum gravity.
\end{abstract}

\maketitle

\section{Introduction}
\label{s1}

The status of loop quantum cosmology (LQC) of FRW models has evolved
significantly over the last year. Specifically, when there is at
least one massless scalar field present, the \emph{physical} sector
of the theory was constructed in detail and then used to show that
the big bang is replaced by a quantum bounce. Furthermore, this
singularity resolution does not come at the cost of introducing
undesirable features such as unphysical matter or large
quantum effects in physically tame situations. We will begin with a
brief summary of these results particularly because there is some
confusion in the literature on the bounce scenario and on the
difference between LQC and the Wheeler DeWitt (WDW) theory.

In these LQC models, the availability of the physical inner product
on the space of solutions to the Hamiltonian constraint and of a
complete set of convenient Dirac observables led to a precise
construction of suitable semi-classical states. Numerical evolution
of these states then led to a number of detailed and quantitative
results: i) Classical general relativity is an excellent
approximation to quantum theory until matter density reaches $\sim
0.01 \rho_{\rm Pl}$, or, scalar curvature reaches $ \sim - 0.15
\pi/\lp^2$; ii) As curvature increases further, quantum geometry
becomes dominant, creating an effective repulsive force which rises
very quickly, overwhelms classical gravitational attraction, and
causes a bounce at%
\footnote{In early papers (e.g., \cite{apslett,aps1,aps2,apsv}) the
bounce was said to occur at $\rho=0.82\rho_{\rm Pl}$. However to
calculate the quantum Hamiltonian constraint these papers used the
lowest non-zero eigenvalue of the area operator for the area gap. It
was later realized that the corresponding eigenstates are not
suitable for homogeneous cosmologies. On the space of states which
\emph{are} suitable, the area gap is twice as large, which makes the
critical density half as large \cite{awe2}.}
$\rho \sim 0.41\rho_{\rm Pl}$, thereby resolving the classical
singularity. The repulsive force dies very quickly once the density
falls below $0.41\rho_{\rm Pl}$; iii) While the classical evolution
breaks down at the singularity, the full quantum evolution remains
well defined, joining the pre-big-bang branch of the universe to the
post-big-bang branch through a \emph{deterministic} evolution; and,
iv) Contrary to the earlier belief, the so-called `inverse volume
effects' associated with the matter Hamiltonian are inessential to
the singularity resolution in these models. At the Planck scale,
dynamics of semi-classical states is dictated, rather, by the
quantum geometry effects in the \emph{gravitational part} of the
Hamiltonian constraint. (For details, see, e.g.,
\cite{apslett,aps1,aps2,apsv,warsaw,aarev}.)

These results also addressed two concrete criticisms \cite{bt,gu} of
the earlier status of LQC. First, although it had been demonstrated
\cite{mb1} that the quantum Hamiltonian constraint remains
well-defined at the putative singularity, in absence of a physical
inner product and Dirac observables the physical meaning of this
singularity resolution had remained unclear. One could therefore ask
\cite{bt}: What is the precise and physically relevant sense in
which the singularity is resolved? Second, while the early
formulations of the quantum Hamiltonian constraint \cite{mb1,abl}
were an improvement over the Wheeler-DeWitt (WDW) theory near the
singularity, they turned out to have a serious flaw: they could lead
to a significant deviation from classical general relativity even in
regimes in which the space-time curvature is quite low \cite{aps1}.
In particular, in the $k$=0 model, more the state is semi-classical,
lower was the matter density and curvature at which the bounce
occurred \cite{aps1,aps2}. In particular, for perfectly reasonable
semi-classical states, the bounce could occur even at density of
water! In presence of a cosmological constant, serious deviations
from classical general relativity also occurred in low curvature
regions well away from the singularity. These problems can also
occur without a cosmological constant. As a result, in the $k$=1
model it was then natural to ask \cite{gu}: does LQC predict that
there would be a recollapse from the expanding to the contracting
phase that can occur at low curvatures in classical general
relativity? Absence of recollapse would be an unacceptable,
\emph{qualitative} departure from general relativity.

Because the physical sector of the theory has now been constructed
in detail, both these issues could be addressed satisfactorily
\cite{aps2,apsv,warsaw}. First, `singularity resolution' refers to
the behavior of Dirac observables such as matter density, whence the
physical meaning of the term is now transparent. Second, not only is
there a recollapse in the $k$=1 model but there is excellent
\emph{quantitative} agreement with general relativity (for universes
which grow to macroscopic sizes) \cite{apsv}. These investigations
have also shown that, while the singularity can be removed rather
easily in LQC, considerable care is needed in the formulation of
quantum dynamics to ensure that the detailed predictions do not lead
to gross departures from classical general relativity in tame
situations \cite{aps1,aps2}. Careful treatment ensures both a good
ultra-violet behavior (singularity resolution) \emph{and} a good
infra-red behavior (agreement with general relativity at low
curvatures) \cite{aps2,apsv}. In particular, in the `improved' LQC
dynamics \cite{aps2,apsv}, the quantum bounce occurs only when the
matter density  is $\sim 0.41 \rho_{\rm Pl}$, irrespective of the
choice of semi-classical states. Once the curvature is low,
evolution is close to that predicted by classical general relativity
also in the presence of a cosmological constant \cite{aps2}.%
\footnote{These important differences between the improved dynamics
(sometimes referred to as the $\bar\mu$ evolution) and the older one
(referred to as the $\mu_o$ evolution) appear not to have been fully
appreciated. Indeed, some of the recent discussions \cite{bck,kks}
portray the two schemes as if they are on equal footing. As
discussed in section \ref{s6}, the issue of whether any of the
proposed dynamics of LQC can be systematically derived from LQG is
wide open but it is logically distinct from whether they are viable
within the confines of LQC. Mixing of these two issues has resulted
in some unfortunate confusion in the literature.}

However, even within the confines of these simple models, a number
of questions still remain. First, numerical simulations and
effective equations both imply that the bounce occurs when the
matter density reaches a critical value $\rcr \approx 0.41 \rho_{\rm
pl}$. It turned out that $\rcr$ is insensitive to the sign or value
of the cosmological constant and is the same whether we consider $k$=0
or $k$=$\pm 1$ models \cite{aps2,apsv,kv}. Can one understand the
physical origin of $\rcr$? Is there an upper bound on matter density
in physical quantum states which can be established without having
to make the assumptions that underlie numerical simulations and
effective equations? Second, the detailed evolution was restricted
to states which are semi-classical at late times and was carried out
numerically. Therefore, an analytical understanding of quantum
evolution has been lacking and, in particular, the precise reason
for the striking differences between LQC and the \WDW theory has not
been fully understood. While states which are semi-classical at late
times are the most interesting ones, are they essential for the
bounce scenario as some authors have suggested? Or, is there a sense
in which all states undergo a quantum bounce? It has been suggested
\cite{mb2} that the near symmetry (in intrinsic time) of volume
uncertainties around the bounce point is a consequence of the use of
a special class of semi-classical states at late times. Is this the
case or is it shared by a much more general class of states which
are (truly) semi-classical at late times?

Similarly, comparison of LQC and the \WDW theory raises a number of
questions. There is a precise sense \cite{abl,aps1,aps2} in which
the quantum Hamiltonian constraint of the \WDW theory approximates
that of LQC. Yet, the physical content of the \emph{solutions} to
the two constraints is dramatically different: while none of the
states of the \WDW theory which are semi-classical at late times can
escape the big bang singularity, all their analogs in LQC do just
that. What then is the precise relation between the two sets of
dynamics? Can one make statements for more general states? While the
geometric part of the \WDW Hamiltonian constraint is a second order
differential operator, that of LQC is a second order difference
operator, the step size being dictated by the `area gap' $\Delta$,
i.e., the smallest non-zero eigenvalue of the area operator. What
happens in the mathematical limit in which $\Delta$ is sent to zero
by hand? Non-relativistic quantum mechanical systems also admit a
`polymer' representation (of the Weyl algebra) which mimics the
mathematical structure of loop quantum cosmology \cite{afw}. For a
harmonic oscillator, dynamics in this representation involves a
(mathematical) discreteness parameter $\ell$ and is distinct from
the standard Schr\"odinger dynamics. However, as was shown in detail
in \cite{afw,cvz,cvz2}, it reduces to the standard dynamics in the
limit in which $\ell$ goes to zero. Is the situation similar in LQC?
If so, what is the a precise sense in which the LQC dynamics reduces
to that of \WDW theory in such a limit?

The purpose of this (and the accompanying \cite{cs}) paper is to
address these questions. Two key ideas will make this task feasible.
First, we introduce a new representation in both LQC and the \WDW
theory in which the operator conjugate to volume is diagonal. In
this representation, the Hamiltonian constraint of LQC becomes a
\emph{differential} operator just as in the \WDW theory (for the
same reasons as in \cite{afw,cvz,cvz2}). Second, we will introduce a
harmonic time coordinate ---which is tailored to the scalar field
clock--- already at the classical level. This will simplify the
factor ordering of the Hamiltonian constraint relative to
\cite{aps2} and make the model exactly soluble. In fact, in the \WDW
as well as in this soluble LQC theory the Hamiltonian constraint
reduces just to the 2-dimensional Klein-Gordon operator! The
physical Hilbert space of both theories is then identical (except
for a certain global symmetry). How can the theories then lead to
strikingly different results on resolution of singularity? The
answer of course is that the (Dirac) observables are represented by
distinct operators. Thus, in the new representation, one can compare
the two theories just by studying the relation between two sets of
operators on the \emph{same} Hilbert space. Furthermore, in this
representation the expression of Dirac observables is rather simple.
Therefore, one can analytically compute the expectation values and
dispersions. We find that in the \WDW theory the expectation value
of the volume operator on a \emph{dense set of states} goes to zero
in the distant past (or future). Thus, for a generic state matter
density diverges in the distant past (or distant future). In this
sense the singularity is unavoidable in the \WDW theory. In LQC by
contrast, on a dense sub-space the expectation value of the volume
operator has a non-zero minimum and diverges both in the distant
past and future. Thus, the density remains finite and undergoes a
bounce. In this sense the quantum bounce is generic and not tied
just to
semi-classical states.%
\footnote{The accompanying paper \cite{cs} analyzes dispersions of
the volume operator in this solvable LQC.}
The simplified model also enables us to show that matter density has
a finite upper bound $\rho_{\rm sup} = \sqrt{3}/ 32\pi^2 \gamma^3
G^2\hbar \approx 0.41 \rho_{\rm Pl}$ on the physical Hilbert space
(where $\gamma$ is the Barbero-Immirzi parameter of LQG). This value
coincides exactly with that of the critical density $\rcr$
\cite{aps2,apsv}! Since the bound is established analytically and
without restriction to states which are semi-classical at late
times, it provides a deeper understanding of the bounce scenario.

Finally, we analyze the precise sense in which the \WDW theory can
be regarded as the continuum limit of LQC. We show that given any
\emph{fixed, finite} interval $I$ of internal time we can shrink (by
hand) the area gap $\Delta$ sufficiently so that the \WDW theory
agrees with LQC to any pre-specified degree of accuracy. However,
this approximation is not uniform in $I$; the \WDW theory cannot
approximate LQC for \emph{all} times no matter how much we shrink
the area gap. Furthermore, LQC does not admit a limit at all if the
gap is shrunk to zero size. In this sense it is intrinsically
discrete, very different from the `polymer quantum mechanics' of a
harmonic oscillator \cite{afw,cvz,cvz2}. We wish to emphasize,
however, that all these results refer only to FRW models with a
massless scalar field.

We conclude this section with a general observation on singularity
resolution. Recall that in classical general relativity, it is
non-trivial to obtain necessary and sufficient conditions to
characterize the occurrence of a singularity (for example, a
space-time can be singular even if all its curvature invariants
vanish). It is therefore not surprising that a generally applicable
and satisfactory notion of `singularity resolution' is not available
in quantum gravity. However, in simple situations such as
homogeneous cosmologies the notion of a singularity is unambiguous
in the classical theory. In the quantum theory of these models one
can also provide a satisfactory notion of singularity resolution.
First, one should have available the physical Hilbert space (not
just the solutions to constraints) and a complete family of Dirac
observables, at least some of which diverge at the singularity. Then
one would say that the singularity is resolved if the expectation
values of these observables remain finite in the regime in which
they become classically singular. In the model studied in this
paper, we have a well defined physical inner product as well as a
complete set of Dirac observables both in the \WDW theory and LQC.
In the \WDW theory the singularity is not resolved because the
expectation value of the matter density diverges at the putative
classical singularities while in LQC the singularity is resolved
because the expectation values of a complete set of Dirac
observables including the matter density remain finite.%
\footnote{ Sometimes apparently weaker notions of singularity
resolution are discussed. Consider two examples \cite{kks}. One may
be able to show that the wave function vanishes at points of the
classically singular regions of the configuration space. However, if
the \emph{physical} inner product is non-local in this configuration
space ---as the group averaging procedure often implies---
vanishing of the wave function would not imply that the probability
of finding the universe at these configurations is zero. The second
example is that the wave function may become highly non-classical.
This by itself would not mean that the singularity is avoided unless
one can show that the expectation values of a family of Dirac
observables which become classically singular remain finite there.}
\medskip

The paper is organized as follows. In section \ref{s2} we introduce
the new representation in the \WDW theory. Section \ref{s3}
introduces the simplification in the LQC Hamiltonian constraint and
the new representation in which this constraint becomes a
differential operator. Section \ref{s4} compares and contrasts the
\WDW theory and LQC and section \ref{s5} provides an analytical
understanding of two striking features of LQC. In section \ref{s6}
we summarize the results and compares them with analogous
discussions in the literature.

\section{The \WDW theory}
\label{s2}

It is well known that the \WDW theory of the $k$=0 FRW models with a
massless scalar field is simple because with an appropriate choice
of variables the Hamiltonian constraint can be cast as a
2-dimensional Klein-Gordon equation (see, e.g.,
\cite{hawking,aps1}). The purpose of this section is to present this
theory in a form that makes its relation to LQC easier to analyze.
For this, we will need to introduce certain variables and notation
which may appear contrived from the internal viewpoint of the \WDW
theory, but which will facilitate comparison between the two
theories in section \ref{s4}.

To construct a Hamiltonian formulation of the $k$=0 model, one has
to introduce a finite fiducial cell $\V$ and restrict all
integrations to it \cite{as,abl}. It is easiest to fix a fiducial
flat metric $\q_{cd}$  and let $\V$ be a cube of volume
$\mathring{V}$ with respect to $\q_{cd}$. The standard canonically
conjugate variables are $(a,\, p_{(a)})$ for geometry and $(\phi,
p_{(\phi)})$ for the scalar field. Here $a$ is the scale factor
relative to $\q_{cd}$: the physical metric $q_{cd}$ being given by
$q_{cd} = a^2\,\, \q_{cd}$. To compare with LQC, we will make a
series of transformations on the geometrical pair $(a, p_{(a)})$.

First, it is convenient to fix a fiducial co-triad $\o_a^i$ which is
orthonormal with respect to $\q_{cd}$ and consider co-triads
$\omega_a^i$ which are orthonormal with respect to $q_{cd}$ so that
$\omega_a^i = a\, \varepsilon\, \o_a^i$, where $\varepsilon =1$ if
$\omega_a^i$ has the same orientation as the fiducial $\o_a^i$ and
$\varepsilon =-1$ if the orientation is opposite. Second, it is
convenient to replace the scale factor $a$ by a variable $\nu$ which
is proportional to volume (of the cell $\V$). Let us set
\be\label{nu1} \nu = \varepsilon\, \f{a^3\, \mathring{V}}{2\pi
\lp^2\gamma}\ee
where $\gamma$ is a constant (the Barbero-Immirzi parameter of
loop quantum gravity) and $\lp = (G \hbar)^{1/2}$ is the Planck length. Thus $\nu$
has the dimension of length and, because of the orientation factor
$\varepsilon$, is not required to be positive; it ranges over
$(-\infty, \infty)$. A canonically conjugate variable is given by
\be \label{b} \b = - \varepsilon\,\f{4\pi \gamma
G}{3\,\mathring{V}}\,\, \f{p_{(a)}}{a^2}\, , \ee
which has dimensions of inverse length. In quantum theory, the
corresponding operators then satisfy the commutation relations:
\be \label{ccr} [\hat{\b},\, \hat{\nu}] = 2i \, . \ee

To construct the Hilbert space of states, we can either use a
representation in which $\hat\nu$ is diagonal or $\hat{\b}$ is
diagonal. In both cases the Hamiltonian constraint becomes a linear,
second order differential operator. Thus both representations are
simple and neither has a particular advantage over the other. In LQC
on the other hand, although the two corresponding representations
are again equivalent, in the $\nu$ representation physical states
turn out to have support on a discrete set of values $\nu =
4n\lambda$,  and the geometric part of the Hamiltonian constraint is
a \emph{difference} operator on states $\Psi(\nu)$ \cite{aps2}. Here
$n$ is an integer and $\lambda$ is the square-root of the smallest
non-zero eigenvalue $\Delta \lp^2$ of area. In the $\b$
representation, on the other hand, states $\Psi(\b)$ have support on
the continuous interval $(0, \pi/\lambda)$ and the geometrical part
of the Hamiltonian constraint is represented by a second order
\emph{differential} operator.%
\footnote{An analogy with a particle moving on a circle makes this
situation transparent. The $\b$ representation in LQC is analogous
to the $\theta$-representation in which the states $\Psi(\theta)$
have support on the interval $(0, 2\pi)$ while the $\nu$
representation of LQC is analogous to the representation in which
$p_\theta$ is diagonal so the the wave-functions $\Psi(p_\theta)$
have support on discrete values $p_\theta = n\hbar$}
Consequently comparison between LQC and the \WDW theory is most
direct in the $\b$ representation. We will therefore let the states
be wave functions $\ul\chi(\b)$ also in the \WDW theory. (As in
\cite{aps1,aps2}, symbols with underbars will refer to the \WDW
theory.)

In this representation, the Hamiltonian constraint takes the form
\be \label{hc1} \partial^2_\phi\,\, \ul\chi (\b,\phi) =  12\pi G
\,\, (\b\,\partial_\b)^2 \,\,\ul\chi(\b,\phi)\, .
 \ee
%
In the classical theory, $\phi$ is monotonic along all dynamical
trajectories and therefore serves as an intrinsic clock. As
explained in \cite{aps1}, because of the form of the Hamiltonian
constraint, this interpretation carries over to quantum theory both
in the \WDW theory as well as LQC. Physical states must satisfy
(\ref{hc1}). Furthermore, they must satisfy a symmetry requirement.
Under change in orientation of the physical co-triad, $(\omega_a^i)
\rightarrow \Pi(\omega_a^i) :=  - \omega_a^i$, the spatial metric
does not change and, since there are no fermions in the model,
physics remains unchanged. Therefore, this change represents a large
gauge transformation. Recall that in Yang-Mills theory physical
states belong to \emph{irreducible} representations of the group of
large gauge transformations (i.e., to the so-called `theta'
sectors). The same reasoning applies to the present case. However,
now the large gauge transformation $\Pi$ satisfies $\Pi^2 = 1$,
whence wave functions
are either symmetric or anti-symmetric under this action.%
\footnote{Detailed considerations imply that physical states
$\ul\chi (\b)$ used here are Fourier transforms of
$\ul{\Psi}(\nu)/\nu$ where $\ul{\Psi}(\nu)$ are the states used in
\cite{aps1,aps2,apsv}. Because of the extra $1/\nu$ factor,
$\ul\chi(\b)$ are anti-symmetric in the \WDW theory (as well as in
LQC). But this fact will not play an essential role in the
subsequent discussion.}
Therefore, it suffices to restrict oneself to just the positive (or
negative) $\b$-half line. For concreteness, let us use the positive
half line.

Then, as in the standard \WDW theory one can simplify the
Hamiltonian constraint by replacing $\b$ with $y$ which ranges over
$(-\infty, \infty)$:
\be \label{y}  y := \f{1}{\sqrt{12\pi G}}\, \ln \f{\b}{\b_o} \quad
\hbox{\rm or, equivalently}\quad \b = \b_o\,\, e^{\sqrt{12\pi G}y}
\ee
where $\b_o$ is a constant of dimensions of inverse length. The
constraint now reduces just to the Klein-Gordon equation in
$(y,\phi)$:
\be \label{hc2} \partial_\phi^2\,\, \ul\chi (y, \phi) =
\partial_y^2\,\, \ul\chi(y,\phi)
=: - \ul\Theta\,\, \ul\chi(y,\phi) .\ee
The physical Hilbert space can be obtained by the group averaging
procedure used in loop quantum gravity \cite{dm,abc,almmt,aps1}. As
usual, the procedure tells us that the physical states can be taken
to be positive frequency solutions to (\ref{hc2}), i.e., solutions
to
\be \label{hc3} -i \partial_\phi\,\, \ul\chi (y,\phi) =
\sqrt{\ul\Theta} \,\, \ul\chi(y,\phi). \ee

Thus, if the initial data at the intrinsic time $\phi=\phi_o$
is\,\,\, $\ul\chi (y,\phi_o) = \f{1}{\sqrt{2\pi}}\,
\int_{-\infty}^\infty \d k\, e^{-iky} \ut{\ul{\chi}}(k)$, \,\, then
the physical state is the solution
\ba \label{sol1} \ul\chi(y,\phi) &=& \f{1}{\sqrt{2\pi}}\,
\int_{-\infty}^\infty
\, \d k \, e^{-iky + i|k|(\phi-\phi_o)}\, \ut{\ul{\chi}}(k) \nonumber\\
 &=& \f{1}{\sqrt{2\pi}}\, \int_{-\infty}^0\, \d k\, e^{-ik(\phi+y)}
 e^{ik\phi_o} \, \ut{\ul{\chi}}(k) +  \f{1}{\sqrt{2\pi}}\, \int_{0}^\infty
 \,\d k \, e^{ik(\phi-y)y}\, e^{-ik\phi_o}\, \ut{\ul{\chi}}(k)\nonumber\\
 &=:& \ul\chi_L(y_+) + \ul\chi_R(y_{-})\,,\ea
where $y_\pm = \phi\pm y$ and the subscripts $L$ and $R$ denote left
and right moving states. The group averaging procedure also implies
that the scalar product is the standard one from Klein-Gordon
theory:
\ba \label{ip1} (\ul\chi_1,\, \ul\chi_2)_{\rm phy} &=& -i
\int_{\phi=\phi_o}\!\!\!\! \d y
\,\,\left[\bar{\ul\chi}_1(y,\phi)\,\,
\partial_\phi\,\, \ul\chi_2(y,\phi) - (\partial_\phi\,\,
\bar{\ul\chi}_1(y, \phi))\,\, \ul\chi_2(y,\phi)
\right]\nonumber\\
&=& 2\int_{-\infty}^{\infty}\! \d k\, |k|\,
\bar{\ut{\ul{\chi}}}_1(k)\, {\ut{\ul{\chi}}_2(k)}\, .\ea
We will denote the resulting physical Hilbert space by $\Hpwdw$. It
is obvious from (\ref{sol1}) and (\ref{ip1}) that the left and right
moving sectors of $\Hpwdw$ are mutually orthogonal. We will see in
section \ref{s4} that physically the left moving component
$\ul\chi_L$ corresponds to the expanding branch of FRW space-times
while and the right moving component $\ul\chi_R$, to the contracting
branch. (Had we worked with the $\b<0$ half line the correspondence
would have reversed.) It is useful to note that the inner product
can also be written as:
\be \label{ip2} (\ul{\chi}_1,\, \ul{\chi}_2)_{\rm phy} =
2\int_{-\infty}^{\infty} \d y \,\, \bar{\ul\chi_1} (y,\phi_o) \,\,\,
|i\p_y|\,\, \,\ul\chi_2 (y,\phi_o) \ee
where the absolute value denotes the positive part of the
self-adjoint operator $i\p_y$.

Our next task is to define a convenient set of Dirac observables.
Since in the classical theory $V|_{\phi_o}$, the volume (of the
fiducial cell) at any instant $\phi_o$ of internal time, is a Dirac
observable, in the quantum theory we wish to define a self-adjoint
operator $\hat{V}|_{\phi_o}$ on $\Hpwdw$. With this goal in mind,
let us first introduce the Schr\"odinger Hilbert space $\H_{\rm
sch}$ by freezing physical states at $\phi=\phi_o$ and evaluating
the scalar product (\ref{ip2}) at $\phi=\phi_o$, and define a
self-adjoint operator $\hat\nu$ on it. Since $\nu$ and $\b$ are
canonically conjugate, we are led to define
\ba \hat\nu &:=& {\hbox{\rm Self-adjoint part of}}\,\,\,
-2i\p_\b\nonumber\\
&\equiv& {\hbox{\rm Self-adjoint part of}}\,\,\, \f{-2i}{\sqrt{12\pi
G}\, \b_o}\, e^{-\sqrt{12\pi G} y} \p_y \, .\ea
Using the fact that the operator $i\p_y$ is a positive definite
self-adjoint operator on the right sector $\H^R_{\rm sch}$ of
$\H_{\rm sch}$ and negative-definite on the left sector $\H^L_{\rm
sch}$, one can easily show that for all smooth functions $\ul\chi_1$
and $\ul\chi_2$ which fall off sufficiently fast at $y=\pm \infty$,
we have
\ba ({\ul{\chi}}_1, -2i\, \p_\b\, {\ul{\chi}}_2)_{\rm sch} =
-\f{4}{\sqrt{12\pi G}\,\b_o}\!\!\!\!\!&&\!\!\!\!\!\!
\int_{-\infty}^\infty\!\! \d y\,\,
\Big[(\overline{i\p_y\ul{\chi}_1^R})\,e^{-\sqrt{12\pi G}y}\,
(i\p_y\ul{\chi}_2^R)\, -\,
(\overline{i\p_y\ul{\chi}_1^L})\,e^{-\sqrt{12\pi G}y}\,
(i\p_y\ul{\chi}_2^L)\, \nonumber\\
&+&(\overline{i\p_y\ul{\chi}_2^R})\,e^{-\sqrt{12\pi G}y}\,
(i\p_y\ul{\chi}_2^L)\, -
(\overline{i\p_y\ul{\chi}_1^L})\,e^{-\sqrt{12\pi G}y}\,
(i\p_y\ul{\chi}_2^R)\,\Big]\nonumber\\ \ea
where the integral is performed at $\phi=\phi_o$. By calculating the
matrix element of the adjoint of $-2i\,\p_\b\,$ and adding we obtain
the matrix elements of $\hat\nu$:
\be ({\ul{\chi}}_1,\, \hat{\nu}{\ul{\chi}}_2)_{\rm sch} =
-\f{4}{\sqrt{12\pi G}\b_o}\, \int_{-\infty}^\infty \d y\,\,
\left[(\overline{i\p_y\ul{\chi}_1^R})\,e^{-\sqrt{12\pi G}y}\,
(i\p_y\ul{\chi}_2^R)\, -\,
(\overline{i\p_y\ul{\chi}_1^L})\,e^{-\sqrt{12\pi G}y}\,
(i\p_y\ul{\chi}_2^L)\right]\ee
Therefore, the desired operator $\hat{\nu}$ on $\H_{\rm sch}$ is
given by
\be \label{nuhat1}\hat\nu = -\f{2}{\b_o \sqrt{12\pi G}}\,\,[ P_R
(e^{\sqrt{12\pi G}\, y}\, i\p_y)P_R\, + \, P_L (e^{\sqrt{12\pi G}\,
y}\, i\p_y)P_L] \ee
where $P_R$ and $P_L$ are projectors on $\H_{\rm sch}^R$ and
$\H_{\rm sch}^L$. We will now use this expression to define a
1-parameter family of (relational) Dirac observables
$\hat{V}|_{\phi_o}$ on $\Hpwdw$:

\be \label{V1} \hat{V}|_{\phi_o}\,\,\ul{\chi}(y,\phi) =
e^{i\sqrt{\ul\Theta}\, (\phi-\phi_o)}\,\,(2\pi \gamma \lp^2\,
|\hat\nu|)\,\, \ul\chi(y,\phi_o)\, .\ee
Thus the action of $\hat{V}|_{\phi_o}$ is obtained by first freezing
the positive frequency solution $\ul\chi (y,\phi)$ at $\phi=\phi_o$,
acting on it by the volume operator $2\pi\gamma \lp^2\, |\hat\nu|$
and evolving the resulting function of $y$ using (\ref{hc3}). (The
operator $|\hat\nu|$ is the positive part of $\hat\nu$ on $\H_{\rm
sch}$.) $\hat{V}|_{\phi_o}$ is a well-defined, self-adjoint operator
on $\Hpwdw$ because $\hat\nu$ enjoys these properties on $\H_{\rm
sch}$.

The second Dirac observable is much simpler: the momentum
$\hat{p}_{(\phi)} = -i\hbar \partial_\phi$. Since it is a constant
of motion, it is obvious that it preserves the space of solutions to
(\ref{hc3}) and is self-adjoint on $\Hpwdw$. Finally, we note that
$\hat{V}|_\phi$ and $\hat{p}_{(\phi)}$ preserves each of the left
and right moving sectors. Since these constitute a complete set of
Dirac observables on $\Hpwdw$, there is superselection and one can
analyze physics of each of these sectors \emph{separately}.

Let us summarize by focusing on the left moving sector for
concreteness. In the more intuitive Schr\"odinger representation,
physical states are functions $\chi_L(y)$ whose Fourier transform
$\tilde{\ul{\chi}}_L(k)$ has support on the negative half of the
$k$-axis and which have finite norm (\ref{ip2}). The matrix elements
of the basic operators are given by
 \ba\label{ops1}
(\ul{\chi}_L,\, \hat{p}_{(\phi)}\, \ul{\chi}_L^\prime)_{\rm phy} &=&
2\hbar\,\int_{-\infty}^{\infty}\!\d y\,\,
\p_y{\bar{\ul\chi}}_L(y,\phi_o)\,
\p_y\ul{\chi}^\prime_L (y, \phi_o) \quad{\rm and} \nonumber\\
(\ul{\chi}_L,   \,\,\hat\nu|_{\phi_o}\,\, \ul{\chi}^\prime_L)_{\rm
phy} &=& \f{4}{\sqrt{12\pi G}\,\, \b_o}\,\,
\int_{-\infty}^{\infty}\! \d y \,\, \p_y \ul{\chi}_L(y,\phi_o)\,\,
e^{-\sqrt{12\pi G}\,y}\,\, \p_y \, \ul{\chi}^\prime_L(y,\phi_o)\,
.\ea
where $\phi_o$ is an arbitrary instant of internal time. Note that
$\hat\nu$ leaves the left sector invariant and is positive definite
on it. Hence, $(\ul{\chi}_L,   \,\,\hat\nu|_{\phi_o}\,\,
\ul{\chi}^\prime_L)_{\rm phy} = (\ul{\chi}_L,
\,\,|\hat\nu|_{\phi_o}|\,\, \ul{\chi}^\prime_L)_{\rm phy}$. Finally,
dynamics is governed by the Schr\"odinger equation (\ref{hc3}).

\medskip

\emph{Remark:} In the $\b$ representation, the entire theory can be
constructed without any reference to the constant $\b_o$. In the $y$
representation by contrast the constant $\b_o$ appears even in the
final theory through the expression of the operator $\hat\nu$. (As
we just saw, the inner product and the expression of the other Dirac
observable $\hat{p}_{(\phi)}$ is independent of $\b_o$ also in the
$y$-representation.) However, it is easy to verify that change in
$\b_o$ just yields a unitarily equivalent theory (as it must). To
make this explicit, denote by $y'$ the left side of (\ref{y})
obtained by replacing $\b_o$ with $\b_o'$. Let us again restrict
ourselves to the left moving sector. Then the theory based on $\b_o$
is mapped to that based on $\b_o' = \alpha \b_o$ by the operator
$\hat{U}_\alpha$:\,\, $\hat{U}_\alpha \chi(y) =\chi'(y') :=
\chi(y'+\ln \alpha/\sqrt{12\pi G})$. This is an unitary map which
preserves $\hat{p}_{(\phi)}$ and sends $\hat\nu$ to $\hat{\nu}'$,
i.e., has the action
\be \hat{U}\, \left[\f{2i}{\sqrt{12\pi G}\, \b_o}\,\, {\exp\,
(-\sqrt{12\pi G}\,y)}\,\, \partial_y \, \right]\,\, \hat{U}^{-1} \,
=\, \f{2i}{\sqrt{12\pi G}\, \b^\prime_o}\,\, {\exp\, (-\sqrt{12\pi
G}\,y^\prime)}\,\,
\partial_{y^\prime} \, .\ee

\section{Solvable LQC}
\label{s3}

This section is divided into three parts. In the first, for
convenience of the reader we present a brief summary of LQC
including a short explanation of the necessity of `improved'
dynamics of \cite{aps2,apsv}. These references began with the
Hamiltonian constraint corresponding to proper time in the classical
theory and changed to the internal time defined by the scalar field
only after quantization. In the second sub-section we follow an
alternate strategy: we use a `harmonic time coordinate' (tailored to
the use of the scalar field as an internal clock) already in the
classical theory. Then the quantum Hamiltonian constraint acquires a
slightly different factor ordering. But this difference turns out to
suffice to enable one to solve the quantum theory exactly. (The
precise relation between the two strategies is discussed in the
Appendix \ref{a1}.) In the third subsection we introduce the $\b$
representation in which the LQC Hamiltonian constraint becomes a
differential operator, facilitating comparison with the \WDW theory
in section \ref{s4}.

\subsection{Dynamics of LQC}
\label{s3.1}

As in the \WDW theory, because of spatial homogeneity and
non-compactness of the spatial manifold, to construct the
Hamiltonian formulation one has to introduce an elementary cell $\V$
and restrict all integrations to it. (This is necessary also in the
path integral treatment based on Lagrangians.) Let us then introduce
fiducial co-triads $\o_a^i$ and triads $\e^a_i$ which define a flat
metric $\q_{ab}$ and choose $\V$ to be a cubical cell whose sides
are aligned with the three $e^a_i$. Symmetries imply that, by a
suitable gauge fixing, the basic canonical pair $(A_a^i, E^a_i)$ of
loop quantum gravity (LQG) can be chosen to have the form \cite{abl}
\be A_a^i = c\,\,\mathring{V}^{-1/3}\,\, \o_a^i, \quad {\rm and}
\quad E^a_i = p\, \sqrt{\det{\q}}\,\, \mathring{V}^{-2/3}\,\, \e^a_i
\ee
where, as before, $\mathring{V}$ is the volume of $\V$ with respect
to the fiducial metric $\q_{ab}$. The dynamical variables are thus
just $c$ and $p$. The factors of $\mathring{V}$ are chosen so that
the Poisson brackets between them are independent of $\mathring{V}$:
\be \{ c,\, p\} = \frac{8\pi\gamma G}{3} \, .\ee
Note that $p$ ranges over the whole real line, being positive when
the physical triad $e^a_i$ has the same orientation as the fiducial
$\e^a_i$ and negative when the orientations are opposite. $c$ is
dimensionless while $p$ has dimensions of area. The two are related
to the scale factor via $p = \varepsilon\, a^2$ (where as before
$\varepsilon= \pm 1$ is the orientation factor), and  $c = \gamma
\dot{a}$ (on classical solutions).

In the LQC Hilbert space, the eigenbasis $\ket{p}$ of $\hat{p}$ is
adapted to quantum geometry. The \emph{physical} volume of the
elementary cell $\V$ is simply $|p|^{3/2}$, i.e. $\hat{V}\ket{p} =
|p|^{3/2} \ket{p}$. The geometrical part of the Hamiltonian
constraint in the connection variables involves the (density
weighted) physical triads $E^a_i$ and the field strength $F_{ab}^i$
of the gravitational connection $A_a^i$. The factors involving
$E^a_i$ can be quantized \cite{abl} using techniques introduced by
Thiemann \cite{tt,ttbook} in full LQG. The field strength $F_{ab}^i$
is more subtle because a fundamental feature of LQG is that while
holonomy operators are well-defined, there is no operator directly
corresponding to the connection \cite{ai,lost}. Therefore, as in
gauge theories, components of $F_{ab}^i$ have to be recovered by
considering holonomies around suitable loops, dividing them by the
area they enclose, and then shrinking these loops.

In the earlier LQC treatments \cite{abl}, this area was calculated
using the \emph{fiducial} metric. The geometrical part of the
Hamiltonian constraint was then a difference operator with uniform
steps in $p$, the step size being dictated by the minimum non-zero
area eigenvalue $2\sqrt{3} \pi \gamma \lp^2 \equiv \Delta \lp^2$.
The resulting dynamics resolves the singularity, replacing the big
bang by a big bounce. However, detailed investigation \cite{aps1}
showed that this dynamics has several unphysical features. First,
for semi-classical states, the density at which the bounce occurs
depends on the expectation value $\hat{p}_{(\phi)}$ in the state,
$\rho_{\rm bounce} = (1/18\pi G \gamma^2)^{3/2}\,
(1/\sqrt{2}|\pphi|)$. Now, $\hat{p}_{(\phi)}$ is a constant of
motion and the higher the value of $\pphi$ the more semi-classical the state
is. (For example, in the closed, $k$=1 model the value of $p_\phi$
determines the maximum volume of the universe $V_{\mathrm{max}}$.
The larger the value of $p_\phi$, the larger is $V_{\mathrm{max}}$). Thus,
more semi-classical the state, lower is the density at which the
bounce occurs, whence the theory predicts that there can be gross
departures from classical general relativity even at density of
water! Second, such departures are not confined just to the bounce.
In presence of a cosmological constant $\Lambda$ they occur also in
a `tame' region well away from the classical singularity, where
$a^2\,\Lambda \equiv |p|\, \Lambda\, \gtrsim 1$. Finally, even
qualitative features of the final quantum dynamics can depend on the
choice of the initial fiducial cell $\V$ (see Appendix B.2 in
\cite{aps2}). The cell is just an auxiliary structure needed in the
mathematical framework. Therefore, while it is acceptable that
intermediate steps in the calculation depend on this `gauge choice',
the final physical predictions must not.

The `improved dynamics' of \cite{aps2,apsv} is based on the
realization that all these limitations stem from the fact that
\emph{fiducial} geometry was used in quantization of $F_{ab}^i$. If
area in this calculation refers instead to the \emph{physical}
geometry, the subsequent mathematics realigns itself just in the
right way to cure all these problems. The density $\rcr$ at the
bounce is now a constant, $\rcr= \sqrt{3}/ (32\pi^2\gamma^3G^2\hbar)
\approx 0.41 \rho_{\rm Pl}$, which is independent of $\pphi$ and
therefore has no implicit dependence on the choice of the fiducial
cell $\V$. In presence of a cosmological constant, departures from
general relativity occur only if the curvature enters the Planck
regime; general relativity continues to be an excellent
approximation away from the singularity, improving steadily as the
scale factor $a$ increases. Thus, the limitations of the older
dynamics stem from the fact that it uses a mathematically variable
but physically incorrect notion of area in defining the operator
analog of $F_{ab}^i$. Indeed quanta of area ---i.e. eigenvalues of
the area operator---  should refer to the physical geometry not to a
kinematical background.

\subsection{The Hamiltonian constraint}
\label{s3.2}

It turns out that the geometrical part of the `improved' Hamiltonian
constraint \cite{aps2} is a difference operator with uniform step
size, but now in volume ($\sim a^3$) rather than $p$ ($\sim a^2$).
Therefore, it is convenient to use a representation in which states
are wave functions $\tilde{\Psi}(\nu)$, where $\nu$ is given by
(\ref{nu1}), so that the operator $\hat{V}$ measuring the volume of
the elementary cell $\V$ is given by%
\footnote{In \cite{aps1,aps2,apsv}, states were chosen to be
functions of a dimensionless parameter $v$ in terms of which volume
eigenvalues are $(8\pi\gamma/6)^{3/2}\, (|v|/K)\, \lp^3$ where $K =
2\sqrt{2}/3\sqrt{3\sqrt{3}}$. Thus, $v$ is related to the present
$\nu$ by $v = 2/\sqrt{8\pi\gamma\sqrt{3}}\, (\nu/\lp)$. This is a
trivial change of variables to simplify the expression of the
Hamiltonian constraint.}
\be \hat{V} \tilde{\Psi}(\nu) = 2\pi\lp^2 \gamma\,
|\nu|\,\tilde{\Psi}(\nu)\, . \ee
As before, the variable $\b$ of (\ref{b}) is canonically conjugate
to $\nu$. However, whereas $\hat{\nu}$ is a well-defined operator,
$\hat{\b}$ is no longer well-defined in LQC because now there is no
operator corresponding to the connection. Nevertheless, since the
holonomies are well-defined, $\widehat{\exp i\lambda\b}$ is a
well-defined operator, with action
\be \label{hol} \widehat{\exp i\lambda\b}\, \tilde{\Psi}(\nu) =
\tilde{\Psi}(\nu+2\lambda) \ee
where $\lambda$ is an arbitrary parameter with dimensions of length.

Since we wish to use the scalar field $\phi$ as an internal time
variable and since $\phi$ satisfies the wave equation $\Box \phi
=0$, it is natural to use a harmonic gauge in which the time
variable $\tau$ satisfies $\Box \tau =0$. The corresponding
space-time metric takes the form:
\be \d s^2 = a^6\, \d\tau^2 + a^2\, \d \vec{x}^2  \ee
where $a\equiv a(\tau)$ is the scale factor. (It is easy to verify
that $\tau$ automatically satisfies $\Box \tau =0$ with respect to
this metric.) Because the lapse function is now given by $N = a^3$,
the Hamiltonian constraint becomes
\be p_{(\phi)}^2 - \f{3}{4\pi G\gamma^2}\, p^2 c^2 =0\, . \ee
The key simplification is the absence of inverse powers of the scale
factor.%
\footnote{This strategy was not used in \cite{aps2,apsv} because
this choice, $N = a^3$ does not have a natural analog outside
homogeneous cosmologies while the choice $N=1$ used there (which
corresponds to using of proper time) is generally available.}
This simplification is sufficient to make the model analytically
solvable. \emph{Therefore, we will call the resulting model}
solvable LQC \emph{and denote it by} sLQC. This labeling will serve
to distinguish the factor ordering used here from that used in
\cite{aps2,apsv}.

The ``improved dynamics'' procedure used in \cite{aps2} now leads to
the quantum constraint:
\be \label{hc4}
\partial_\phi^2\, \tilde{\Psi}(\nu,\phi) = 3\pi G\, |\nu|\,
\f{\sin\lambda\b}{\lambda}\, |\nu|\, \f{\sin\lambda\b}{\lambda}\,
\tilde{\Psi}(\nu,\phi)\, \ee
where $\lambda$ is now the (positive) square-root of the area gap:
$\lambda^2 = \Delta \lp^2 \equiv 2\sqrt{3}\, \pi\gamma\lp^2$. Since
one of the goals of this paper is to spell out the relation between
LQC and the \WDW theory, we want to retain the ability of sending
the area gap to zero. Therefore \emph{we will leave $\lambda$ as a
free parameter, $\lambda \le \sqrt{\Delta}\, \lp$, rather than
fixing it to the LQC value $\lambda = \sqrt{\Delta}\, \lp$.}

Next, as in the \WDW theory, physical states have to satisfy a
symmetry requirement. Under orientation reversal $\Pi$ of physical
triads $e^a_i$, $\nu \rightarrow \Pi(\nu) = -\nu$. This is a large
gauge transformation. As discussed in section \ref{s2}, following
the standard procedure from gauge theories, physical states belong
to irreducible representations of the group of large gauge
transformations. Since there are no fermions in this model, in LQC
$\tilde{\Psi}$ is assumed to be symmetric: $\Pi
\tilde{\Psi}(\nu,\phi) := \tilde{\Psi}(-\nu,\phi) =
\tilde{\Psi}(\nu,\phi)$.

Using this symmetry and writing out the explicit action of operators
$\sin \lambda\b$, (\ref{hc4}) simplifies to:
\ba \label{hc5} \partial_\phi^2 \,\tilde{\Psi} (\nu, \phi) &=& 3\pi
G\, \nu\, \f{\sin\lambda\b}{\lambda}\, \nu\,
\f{\sin\lambda\b}{\lambda}\, \tilde{\Psi}(\nu,\phi) \nonumber\\
&=&\f{3\pi G}{4\lambda^2}\, \nu \left[\, (\nu+2\lambda)
\t\Psi(\nu+4\lambda) - 2\nu \t\Psi(\nu) + (\nu -2\lambda)
\t\Psi(\nu-4\lambda)\, \right]\nonumber\\
&=:& \Theta_{(\nu)}\, \t\Psi(\nu,\phi)\, . \ea
The geometrical part, $\Theta_{(\nu)}$, of the constraint is a
difference operator in steps of $4\lambda$. Hence, as discussed in
\cite{aps2,apsv}, there is again superselection: for each $\epsilon
\in [0,4\lambda)$, the space of states $\Psi(\nu)$ with support on
points  $\nu = \epsilon + 4n\lambda$ is preserved under dynamics. In
this paper we will focus on the $\epsilon =0$ `lattice' which is
invariant under $\Pi$.

\subsection{Reducing the sLQC constraint to a Klein-Gordon Equation}
\label{s3.3}

For reasons explained in the Introduction, we now wish to work in
the $\b$ representation because the geometrical part of the quantum
constraint will also become a differential operator. Since
$\t\Psi(\nu,\phi)$ have support on the `lattice' $\nu = 4n\lambda$,
and since $\b$ is canonically conjugate to $\nu$, their Fourier
transforms $\Psi(\b,\phi)$ have support on the continuous interval
$(0, \pi/\lambda)$:
\be \Psi(\b,\phi) := \sum_{\nu=4n\lambda}\, e^{\ihalf \nu\b}\,\,
\t\Psi(\nu,\phi); \quad \hbox{\rm so that} \quad \t\Psi(\nu, \phi) =
\f{\lambda}{\pi}\, \int_0^{\pi/\lambda} \!\! \d\b\, e^{- \ihalf
\nu\b}\,\, \Psi(\b,\phi)\, .\ee
{}From the form (\ref{hc4}) of the constraint it is obvious that it
would be a second order differential operator in the
$\b$-representation. To facilitate comparison with the \WDW theory,
let us set $\tilde\chi(\nu,\phi) = (\lambda/\pi
\nu)\tilde{\Psi}(\nu,\phi)$. Then, on $\chi(\b,\phi)$, the
constraint (\ref{hc4}) becomes
\be \label{hc7}
\partial^2_\phi \, {\chi}(\b,\phi) = 12\pi G\,\, \left(\f{\sin
\lambda\b}{\lambda}\, \partial_\b\right)^2\,\, {\chi}(\b,\phi) \ee
which is strikingly similar to the \WDW equation (\ref{hc1}). Note
however, that \emph{we did not} arrive at (\ref{hc7}) simply by
replacing $\b$ in the expression of the classical constraint by
$\sin\lambda\b/\lambda$ as is often done (see, e.g.,
\cite{mb2,other}). Rather, (\ref{hc7}) results directly from the
`improved' LQC constraint if one begins with a harmonic time
coordinate already in the classical theory.

To simplify the constraint further, let us set
\be\label{x}  x = \f{1}{\sqrt{12\pi G}}\, \ln (\tan
\f{\lambda\b}{2}),\quad \hbox{\rm or}\quad \b = \f{2}{\lambda}\,
\tan^{-1}\, (e^{\sqrt{12\pi G}\, x})\ee
so $x$ ranges from $-\infty$ to $\infty$. Then (\ref{hc5}) becomes
just the Klein-Gordon equation
\be \label{hc8}\partial^2_\phi\,\, \chi(x,\phi) =
\partial_x^2\,\,\chi(x,\phi) =: -\Theta\,\, \chi(x,\phi)\, .\ee
Therefore we can repeat the discussion of section \ref{s2}. The
physical Hilbert space is again given by positive frequency
solutions to (\ref{hc8}), i.e. satisfy
\be \label{hc9} -i \p_\phi \chi(x,\phi) = \sqrt{\Theta}\, \chi
(x,\phi)\, . \ee
We can again express the solutions in terms of their initial data
and decompose them into left and right moving modes $\chi(x,\phi)=
\chi_L(x_+)+ \chi_R(x_-)$ (see Eq (\ref{sol1})). The physical inner
product is given by (\ref{ip1}) or, equivalently, (\ref{ip2}). Thus,
the physical Hilbert space of LQC is exactly the same as in the \WDW
theory and indeed the action of the operator $\hat{p}_{(\phi)}$ on
physical states is the same: $\hat{p}_{(\phi)} \chi = -i\hbar
\p_\phi \chi \equiv \sqrt{-\p_x^2} \chi$.

How can there be difference in physical predictions, then? Recall
that to obtain physical predictions, we need the action of a
complete set of physical observables and, as we will now show, the
volume observable $\hat{V}|_{\phi_o}$ in LQC has a \emph{different}
form from that in the \WDW theory. Consequently, although the
Hilbert spaces are mathematically the same, the \emph{physical
meaning} of a given positive frequency solution is different in the
two theories.

Let us begin with the operator $\hat\nu$ on the Schr\"odinger Hilbert
space $\H_{\rm sch}$ spanned by the initial data $\chi(x,\phi_o)$ to
(\ref{hc9}), where the inner product is obtained by evaluating
(\ref{ip2}) at $\phi=\phi_o$. As in the \WDW theory, $\hat\nu$ is
again the self-adjoint part of $-2i \partial_b$. \emph{However,
while $-2i\p_b = (-2i/\sqrt{12\pi g}\b_o)\,\exp(-\sqrt{12\pi\, G}\,
y)\, \p_y$ in the \WDW theory, it is now given by
$(-2i\lambda/\sqrt{12\pi G})\, \cosh (\sqrt{12\pi G}\, x)\,\p_x$.}
Therefore, repeating the procedure we followed in section \ref{s2},
on $\H_{\rm sch}$ the operator $\hat\nu$ is given by:
\be \label{nuhat2}\hat\nu = -\f{2\lambda}{\sqrt{12\pi G}}\,\,[ P_R
(\cosh({\sqrt{12\pi G}\, x})\, i\p_x)\,P_R\, + \, P_L
(\cosh({\sqrt{12\pi G}\, x})\, i\p_x)\, P_L]\, . \ee
Therefore, as in the \WDW theory, the corresponding volume operator
on $\H_{\rm phy}$ is given by
\be \label{V2} \hat{V}|_{\phi_o}\,\,{\chi}(x,\phi) =
e^{i\sqrt{\Theta}\, (\phi-\phi_o)}\,\,(2\pi \gamma \lp^2\,
|\hat\nu|)\,\, \chi(x,\phi_o)\, .\ee

A second difference from the \WDW theory arises because of symmetry
conditions on the physical states. Recall that to qualify as a
physical state $\Psi(\nu,\phi)$ has not only to be a positive
frequency solution but must also satisfy $\t\Psi(-\nu,\phi)=
\t\Psi(\nu,\phi)$. This translates to the condition $\chi(-x,\phi) =
-\chi(x,\phi)$. Therefore $\chi(x,\phi)$ has the form
\be \label{symmetry} \chi (x,\phi) = \f{1}{\sqrt{2}}\, (F(x_+) -
F(x_-)) \ee
for some function $F$ (such that $F(x_\pm)$ are positive frequency
solutions to (\ref{hc8})). Consequently, in contrast to the \WDW
theory, the right and left sectors are \emph{not} superselected.
However, since the full information in any physical state
$\chi(x,\phi)$ is contained in $F$, we can conveniently describe the
$\Hp$ in terms of positive frequency, left moving solutions $F(x_+)$
(or, right moving solutions $F(x_-)$) alone, which are free from any
symmetry requirement. In this description, the scalar product is
given simply by
\be \label{ip5} (\chi_1, \chi_2)_{\rm phy} =
-2i\int_{-\infty}^{\infty}\!\! \d x \,\, \bar{F}_1(x_+) \partial_x
F_2(x_+) \ee
where the integral is evaluated at a constant value of $\phi$.

Let us summarize the basic structure using a Schr\"odinger
representation which will be useful in the next section. For any
given $\lambda$, the LQC Hilbert space can be taken to be the space
of functions $F(x)$ whose Fourier transform $\tilde{F}(k)$ has
support only on the positive half line and whose norm given by
(\ref{ip5}) is finite. Matrix elements of the basic observables
$\hat{p}_{(\phi)}$ and $\hat{\nu}$ are given by:
\ba\label{ops2} (F_1,\, \hat{p}_{(\phi)} F_2)_{\rm phy} &=&
2\hbar\,\int_{-\infty}^{\infty}\!\d x\, \p_x \bar{F}_1(x_+)\, \p_x
F_2(x_+) \quad {\rm and}\nonumber\\
(F_1, \,\hat\nu|_{\phi_o}\,\, F_2)_{\rm phy} &=&
\f{4\lambda}{\sqrt{12\pi G}}\,\, \int_{-\infty}^{\infty}\! \d
x\,\p_x \bar{F}_1(x_+)\,\, \cosh (\co x)\,\, \p_x  F_2(x_+)\ea
where the integral is evaluated at $\phi=\phi_o$. The states evolve
via the Schr\"odinger equation
\be -i{\p_\phi} F(x_+) = \sqrt{\Theta} F(x_+) \ee
so that, as in the \WDW theory the effective Hamiltonian is
non-local in $x$. Thus, the only difference from the \WDW theory
lies in the expression of $\hat{\nu}$ and hence of the volume
operator.

\medskip

\textit{Remark}: A quick way to arrive at the constraint (\ref{hc7})
in the b-representation is to write the classical constraint in
terms of the canonical pair $(\b,\nu)$ and then simply replace $\b$
by $(\sin\,\lambda\b) /\lambda$ and $\nu$ by $-2i\p_\b$. While this
so-called `polymerization method' \cite{cvz2,other} yields the
correct final result, it is not directly related to procedures used
in LQG. Since $\nu$ is a geometrical variable, its quantization
could be carried out using quantum geometry techniques. However, $\b$
has no natural analog in LQG. In particular, since it is not a connection
component, a priori it is not clear why the wave functions have to
be even almost periodic in $\b$  nor why $\lambda$ should be related
to the area gap.
For a plausible relation to LQG one has to start with the canonical
pair $(A_a^i, E^a_i)$, i.e., $(c,p)$, mimic the procedure used in
LQG as much as possible, e.g., along the lines of \cite{aps2}, and
then pass to the $\b$ representation as was done here. Once this is
done, \emph{a posteriori} it is possible, and indeed often very
useful, to use shortcuts such as $\b \rightarrow
(\sin\lambda\,\b)/\lambda$.

\section{The \WDW theory and LQC: Similarities and differences}
\label{s4}

This section is divided into two parts. In the first we show that
there is a precise sense in which the singularity is generic in the
\WDW theory while it is generically replaced by a quantum bounce in
sLQC. In the second we spell out a precise sense in which the \WDW
theory approximates sLQC and the sense in which this approximation
fails.

\subsection{Singularity versus the quantum bounce}
\label{s4.1}

In spatially homogeneous, isotropic space-times the only curvature
invariant is the space-time scalar curvature which is proportional
to the matter density. Classical singularity is characterized by the
divergence of these quantities. Now, the matter density is given by
$\rho = \pf^2/2V^2$ and since there is no potential for the scalar
field, $\pf$ is a constant of motion in our model (also in the
quantum theory). Therefore, to answer the question of whether a
given quantum state is singular or not, one can calculate the
expectation values of the Dirac observable $\hat{V}_\phi$. If $\vex$
goes to zero (possibly in the limit as $\phi$ tends to $\pm\infty$
as in classical general relativity), one can say that the quantum
state leads to a singularity. If $\vex$ diverges in both directions,
i.e., as $\phi \rightarrow \pm\infty$, but attains a non-zero
minimum, one can say that the state exhibits a quantum-bounce. While
this is a rather weak criterion, it is applicable to any state (in
the domain of the volume operator). In this sense, it enables us to
test the generic behavior in the \WDW theory and in sLQC, thereby
going beyond the analysis of \cite{apslett,aps1,aps2} which was
restricted to states which are semi-classical at late times.

Let us begin with the \WDW theory. As noted in section \ref{s2} the
left and right moving sector are superselected and can be analyzed
separately. For definiteness we will focus on the left moving
sector. Then, for any $\ul\chi(y_+)$ we have
\ba \label{wdwvol}(\ul{\chi}_L,\, \hat V|_{\phi} \,
\ul{\chi}_L)_{\mathrm{phy}} &=& \nonumber 2 \pi \gamma \lp^2 \,
(\ul{\chi}_L,\,\,|\hat \nu|_{\phi_o}\,\ul{\chi}_L)_{\mathrm{phy}} \\
&=& \nonumber  \f{8\pi \gamma \lp^2}{\co \b_o} \,
\int_{-\infty}^{\infty}\, \d y \, \left|\f{\p  \ul{\chi}_L}{\p
y}\right|^2\, e^{-\co y} \\
&=&  \nonumber\left[\,\f{8\pi \gamma \lp^2}{\co\, \b_o} \,
\int_{-\infty}^{\infty} \, \d y_+ \, \left|\f{\d  \ul{\chi}_L}{\d
y_+}\right|^2 \, e^{-\co (y_+)}\, \right]\,\, e^{\co \phi} \\
&=:& V_o \, e^{\co \phi} \, , \ea
where $V_o$ is a constant determined by the solution and can be
calculated from initial data at any instant of time. Thus, for any
state $\ul{\chi}_L(y_+)$ in the domain of the volume operator, the
expectation value $\vex$ tends to $\infty$ as the internal time
$\phi$ tends to $\infty$ and goes to zero as $\phi$ tends to
$-\infty$. In this sense the left moving sector corresponds to the
expanding universes. It is obvious from the above calculation that
on the right moving sector $\ul{\chi}_R(y_-)$ the situation would be
reversed whence it represents contracting universes. In the first
case there is a big bang singularity and in the second, a big crunch
singularity. Indeed, states which are semi-classical in an epoch
when the density is very low compared to the Planck density are
known to follow the classical trajectories into either the big bang
or the big-crunch singularity \cite{aps2}. We have shown that this
qualitative behavior is generic: for a dense subset of states,
expectation values $\vex$ of the volume operator evolve into a big
bang or a big crunch singularity. This calculation also shows that
the matter density, defined as $\tilde\rho := \langle
\hat{p}_{(\phi)} \rangle^2/2 \langle \hat{V} \rangle^2$ diverges as
$\phi \rightarrow -\infty$ (resp. $\phi\rightarrow \infty$) in the
left (resp. right) moving sector. Furthermore, the same
considerations apply to the expectation values of the density
operator $\hat\rho|_{\phi}$
because, as shown in section \ref{s5.1}, its expectation value has
the same behavior as $\tilde\rho$.

Let us carry out the same calculation in sLQC. Consider any state
$\chi(x,\phi) = (1/\sqrt{2})\,(F(x_+) - F(x_-))$. Then, we have:
\ba (\chi,\,\, \hat V|_{\phi} \, \chi)_{\mathrm{phy}}&=&\nonumber
2 \pi \gamma \lp^2 \, (\chi,\,\,|\hat \nu|_{\phi_o}\, \chi)_{\mathrm{phy}}\\
&=& \nonumber \f{8\pi \gamma \lp^2 \lambda}{\co} \,
\int_{-\infty}^{\infty}\!\d x \,\big[\p_x \bar\chi_L\, \cosh(\co\,
x) \, \p_x\chi_L + \p_x \bar\chi_R\, \cosh(\co\, x) \, \p_x\chi_R
\big]
\\
&=& \no \f{8\pi \gamma \lp^2 \lambda}{\co} \,
\int_{-\infty}^{\infty} \! \d x_+ \, \left| \f{\d F}{\d x_+}
\right|^2 \cosh(\co (x_+ - \phi))\, , \ea
which can be written as
\be  \label{lqcvol}  (\chi,\hat V|_{\phi} \, \chi)_\phy = V_+ \,
e^{\co \phi} + V_- \, e^{-\co \phi} \ee
where
\be\label{vpm} V_{\pm} = \f{4 \pi\gamma \lp^2 \lambda}{\co} \,
\bigg[\int_{-\infty}^{\infty} \, \d x_+ \, \left| \f{\d F}{\d x_+}
\right|^2 e^{\mp\co x_+}\bigg]\ee
are again constants associated with the solution which can be
calculated from the initial data at any fixed $\phi$. Note that
$V_\pm$ are strictly positive (compare \cite{mb2}). In contrast to
the \WDW theory, $\vex$ tends to infinity both in the distant future
and in the distant past. Furthermore, it has a \emph{unique} local
minimum which is therefore also a global minimum. In this sense the
bounce picture of \cite{aps2,apsv} is robust. It is not restricted
to states which are semi-classical at late times but holds for all
states in the domain of the volume operator. Similarly, since the
result is obtained analytically, an apparent concern (which arose
from some observations made in \cite{pl}) that the presence of the
bounce may be related to intricacies of numerics is ill founded. The
bounce point and the minimum volume are given by:
\be\label{bounce} \phi_{\B} = \f{1}{2 \co} \, \log \f{V_-}{V_+},
\quad {\rm and} \quad  V_{\rm min} =2 \f{(V_+\,V_-)^{\f{1}{2}}}
{||\chi||^2}\, , \ee
where $||\chi||$ is the norm of the state $\chi$.

Next, we note that
\be \vex_{\phi_{{\B}}+\phi} = V_{\rm min}\, \cosh (\co\,\phi) \, .
\ee
Therefore, the internal-time evolution of volume is exactly
symmetric about the bounce point for \emph{all} states. For reasons
just discussed one would expect that the symmetry would hold also in
LQC to an excellent degree of approximation. This symmetry was already
observed in \cite{aps2} through numerical and effective equation methods and
in \cite{mb2} using a simplified model. However, all
these analyses were tied to various notions of semi-classicality. Our
results overcome the concern that the symmetry may not extend to general
states.
Our
considerations here refer only to expectation values. However, the
same techniques are used in \cite{cs} to calculate fluctuations of
the volume operator and analyze its properties.

\emph{Remark:} In the \WDW theory the left and right sectors are
superselected and should therefore be considered separately.
Nonetheless, since in sLQC all physical states $\chi$ satisfy
$\sqrt{2}\,\chi(x,\phi) = F(x_+) - F(x_-)$ for some $F$, one may be
tempted to consider states $\ul{\chi}(y,\phi)$ which are similarly
anti-symmetric in $y$ also in the \WDW theory. This strategy would
however be physically incorrect because of two reasons. First, the
anti-symmetry condition in $x$ arises in sLQC because physics is
symmetric under reflections of spatial triads $E^a_i$. This
condition has already been incorporated in the \WDW theory in
arriving at the $y$ representation. Secondly, if one were
nonetheless to impose anti-symmetry in the \WDW theory, one would
find that this sector does not have any semi-classical states at
very late (or very early) times because, in this sector, states
which are peaked at large volume would have very large ---rather
than small--- extrinsic curvature.

\subsection{Approximating sLQC with the \WDW theory}
\label{s4.2}

There is a precise sense in which the \WDW equation approximates
the LQC evolution equation \cite{abl,aps1,aps2}. However, as is
well known, the issue of whether solutions to two such equations
approximate each other is logically quite distinct. In particular
the equations may be such that initial data satisfying the
assumptions needed to show that the two equations approximate each
other may evolve out of this regime. If this occurs, solutions of
the two theories may be close to each other initially but could
eventually differ drastically. In this section we will show that
this is precisely what happens in our case.

At first this may seem perplexing because in sections \ref{s2} and
\ref{s3} we showed that the physical states in both theories satisfy
the (positive frequency) Klein-Gordon equation. Isn't then the
evolution identical? However, as pointed out in section \ref{s3.3}
that in quantum theory the physical content of a state lies not in
its functional form but in its interplay with observables. While the
action of $\hat{p}_{(\phi)}$ is the same in the two theories, the
action of $\hat{V}|_\phi$ is distinct (compare (\ref{ops1}) and
(\ref{ops2})). Therefore the theories are inequivalent. The question
for us is: Can one arrive at the \WDW theory by shrinking the area
gap of sLQC, i.e., by letting $\lambda$ go to zero? Again, one
cannot naively set $\lambda$ to zero: $\lambda$ appears linearly in
the expression of the volume operator (see (\ref{ops2})) and setting
it to zero will make the operator itself zero. Rather, we have to
work with states and operators together.

Let us consider the 1-parameter family of Dirac observables
$\hat{V}|_{\phi}$. A \emph{necessary} condition for the \WDW theory
to approximate sLQC is that, for any state of the \WDW theory, there
should exist a state of sLQC${{}_{(\la)}}$ such that the expectation
values of volume operators in the two states can be made arbitrarily
close to one another by choosing a sufficiently small
$\lambda$.%
\footnote{In the $k$=0, spatially non-compact context considered
here, one must introduce a finite cell to speak of volume. However,
ratios of the volume expectation values $\vex$ at two different
times or, at the same time but computed using sLQC and the \WDW
theory are independent of the choice of the cell.}
More precisely, given $\epsilon >0$, there should exist a $\delta
>0$ such that for any state $\ul\chi_L(y,\phi)$ of the \WDW theory,
there is a state $\chi_{{}_{(\la)}}(x,\phi)$ of sLQC${{}_{(\la)}}$
such that
\be \label{lim} |\vex_{(\la)} - \vex_{\rm wdw}| < \epsilon \ee
for all $\la < \delta$. Now, the forms (\ref{wdwvol}) and
(\ref{lqcvol}) of the two expectation values
\be \label{evo} \vex_{\rm wdw} = V_o e^{\co \phi}, \quad{\rm and}
\quad \vex_{(\la)} = V_+ e^{\co \phi} + V_- e^{-\co \phi} \ee
immediately implies that if (\ref{lim}) is to hold for all
positive $\phi$ then
\be V_o = V_+ \quad\quad {\rm and} \quad\quad V_- < \epsilon \ee
where $V_o$ is defined by (\ref{wdwvol}) using $\ul\chi_L(y,\phi)$
and $V_\pm$ by (\ref{vpm}) using $\chi_{{}_{(\la)}}(x,\phi)$.
However, since $V_-$ is necessarily positive, it follows that
$|\vex_{(\la)} - \vex_{\rm wdw}|$ will grow unboundedly as $\phi
\rightarrow -\infty$ no matter how small $\delta$ is. Thus, the
necessary condition for the \WDW theory to approximate sLQC is
violated: the global dynamics of the two theories is very different.

While the simple argument above establishes the main result of this
section, it is instructive to examine some details by explicitly
constructing a map from $\ul\chi_L(y,\phi)$ to
$\chi_{{}_{(\la)}}(x,\phi)=(1/\sqrt{2})( F_{(\lambda)}(x_+) -
F_{(\lambda)}(x_-))$. For this it is convenient to work in the
Schr\"odinger representation at time, say, $\phi=0$. Then our task
is to construct $F_{(\lambda)}(x)$ from the initial data
$\ul\chi_L(y,0)$. Fix an $\epsilon >0$ and consider the family of
Schr\"odinger states $\ul\chi_L(y)$ of the \WDW theory for which
$V_-$ of (\ref{vpm}) (obtained by using $\ul\chi_L$ for $F$)
satisfies $V_- <\epsilon$. Then, considerations in the remark at the
end of section \ref{s2} suggest that we set
\be \label{flow} F_{(\la)}(x) := \ul\chi_L(x+\mu)\quad\quad {\rm
where} \quad\quad \mu = \f{1}{\co}\, \ln\, \f{2}{\b_o\lambda}\, .
\ee
Then for any $\la$, the pairs $\ul\chi(y)$ and $F_{(\la)}(x)$ are
Schr\"odinger states which are `close' to one another at $\phi=0$
in the sense that:\\
i) They have the same norm:
$||\ul\chi(y)||_{\phy}^{\rm wdw}
= ||F_{(\la)}(x)||_{\phy}$; and,\\
ii) They have the same expectation values for any power of
$\hat{p}_\phi$: ${\langle \hat{p}_{(\phi)}^n\rangle}_{\rm wdw} =
{\langle\hat{p}_{(\phi)}^n\rangle}_{(\la)}$; and, \\
iii) Their volume expectation values are close: $|{\langle
\hat{V}|_\phi\rangle}_{\rm wdw} - {\langle
\hat{V}|_\phi\rangle}_{(\la)}|  < \epsilon$.\\

Note that since $\mu$ depends on $\lambda$, for each $\lambda$ the
prescription chooses a different initial state in
sLQC${}_{(\lambda)}$. If $\ul\chi(y)$ were peaked at some value
$y_o$, each $F_{(\la)}(x)$ will also be peaked at some $x_o$.
However, as $\lambda$ decreases this peak would shift progressively
to $x= -\infty$. This change of the functional form of
$F_{(\la)}(x)$ with $\lambda$ is the `renormalization flow' required
to ensure that physics of the initial state remains the same at each
scale $\lambda$.

Let us now evolve these states. Since $\hat{p}_{(\phi)}$ is a
constant of motion, the non-trivial dynamics is in the expectation
values of the volume operator. (\ref{evo}) immediately implies that
while the difference between the two sets of expectation values will
in fact shrink as $\phi$ increases, in the distant past it will grow
unboundedly! Thus, the two sets of dynamics approximate each other
well only on a semi infinite time interval (which is finite in the
negative $\phi$ direction).

What happens when we shrink $\lambda$? Since
\be V_{-} = \f{4 \pi\gamma \lp^2 \lambda}{\co} \,
\bigg[\int_{-\infty}^{\infty} \, \d x \, \left| \f{\d F}{\d x}
\right|^2 e^{\co x}\bigg]\, ,\ee
and since $F_{(\la)}(x) = \ul\chi(x+\mu)$ where $\mu$
monotonically decreases and  $\mu \rightarrow -\infty$ as $\lambda
\rightarrow 0$, $V_-$ decreases steadily in this limit. Therefore,
it follows from (\ref{lqcvol}) that the condition $|\vex_{\rm
wdw}(\phi) - \vex_{(\la)} (\phi)| <\epsilon$ is satisfied further
and further in the distant past. Therefore, given an $\epsilon
>0$ and \emph{any} semi-infinite time interval $I = (-\phi_o,
\infty)$, \emph{however large $\phi_o$ may be},  there exists a
$\lambda >0$ such that the dynamical evolution of $\langle \hat{V}
\rangle$ in sLQC${}_{(\lambda)}$ dynamics remains within $\epsilon$
of that in the \WDW theory on $I$. In particular, as observed in
\cite{aps2}, as we shrink $\lambda$, for the same initial data at
$\phi=0$, the bounce time $\phi_{\B}$ is pushed further and further
into the past.

To summarize, the LQC dynamics does \emph{not} reduce to the \WDW
dynamics as the area gap shrinks to zero. However, if one is
interested only in semi-infinite intervals such as $(-\phi_o,
\infty)$, one can recover the \WDW dynamics on choosing \emph{by
hand} a sufficiently small $\lambda$.

\section{Two striking features of sLQC}
\label{s5}

This section is divided into two parts. In the first, we show that
matter density admits an absolute upper bound $\rho_{\rm sup}$ on
the physical Hilbert space which, furthermore, equals the critical
density $\rcr$ found in \cite{aps2,apsv} using effective equations
and numerical evolutions of the exact equations. In the second, we
show that there is a precise sense in which sLQC is a fundamentally
discrete theory.

\subsection{An absolute upper bound on matter density}
\label{s5.1}

The fact that \emph{every} LQC physical state undergoes a bounce at
a positive value of $\langle \hat{V}|_{\phi_{\rm B}}\rangle$
provides a sense in which the singularity is resolved in LQC. We
will now show that there is a much stronger sense in which the
resolution occurs: the spectrum of the density operator
$\hat{\rho}|_\phi$ is bounded above on the full physical Hilbert
space $\Hp$. Note that the boundedness of volume by itself is
\emph{not} sufficient to imply the boundedness of $\hat{\rho}|_\phi$
\emph{because the operator $\hat{p}_{(\phi)}$ does not have a finite
upper bound} on $\Hp$.

In the classical theory, the scalar field density is given by
$\rho|_\phi = (1/2)(p_{(\phi)}/V|_\phi)^2$. Since the operators
$\hat{p}_{(\phi)}$ and $\hat{V}|_\phi$ do not commute in quantum
theory, it is natural to define the density operator as
\be \hat\rho|_\phi = \f{1}{2}\, \hat{A}|_\phi^2\quad {\rm where}
\quad \hat{A}|_\phi = (\hat{V}|_\phi)^{-1/2}\, \hat{p}_{(\phi)}\,
(\hat{V}|_\phi)^{-1/2}\ee
Now, because $\hat{p}_{(\phi)}$ is a constant of motion, in any
given physical state one would expect the time-dependent observable
$\hat\rho|_{\phi}$ to reach its maximum value $\rho_{\B}$ at
internal time $\phi=\phi_{\rm B}$. Numerical simulations have shown
\cite{aps2,apsv} that for states which are semi-classical at late
times the value of $\rho_{\B}$ is remarkably robust: $\rho_{\B}
\approx 0.41 \rho_{\rm Pl}$ for all the states considered in models
with and without a cosmological constant and also for the $k=\pm 1$
cosmologies. (It was therefore called critical density and denoted
$\rcr$). It is natural to ask if this fact has an analytical
explanation in sLQC. We will now show that the answer is in the
affirmative.

Let us first compute the expectation values of the operator
$\hat{A}$ using those of $\hat{p}_{(\phi)}$ and $\hat{\nu}|_{\phi}$
given in  (\ref{ops2})
\ba \langle \hat{A}\rangle &:=& \f{(\Psi,\,
\hat{A}|_{\phi_o}\Psi)_{\rm phy}}{ (\Psi, \Psi)_{\rm phy}}\,\, =\,\,
\f{(\chi,\,\, \hat{p}_{(\phi)}\, \chi)_{\rm phy}}{ (\chi,\,\,
\hat{V}|_{\phi_o}\, \chi)_{\rm phy}}\nonumber\\
&=& \left(\f{3}{4\pi\gamma^2 G}\right)^{\f{1}{2}}\,\, \f{1}{\la}\,\,\,
\f{\left[\int_{-\infty}^{\infty}\! \d x |\p_x F|^2\right]}
{\left[\int_{-\infty}^{\infty}\! \d x |\p_x F|^2\,\, \cosh (\co x)
\right]}\, \ea
where
\be \chi (x,\phi) = \hat{V}^{-\f{1}{2}} \Psi = \f{1}{\sqrt{2}}\,
(F(x_+) - F(x_-)) \ee
and the integrals are performed at $\phi=\phi_o$. Since $\cosh (\co
x) \ge 1$, it follows that the ratio of the the two integrals is
bounded above by 1. Therefore, using the fact that $\lambda^2 =
2\sqrt{3}\gamma\lp^2$ in LQC, we obtain
\be \label{Abound} \langle \hat{A}\rangle \le \left(\f{3}{4\pi\gamma^2
G}\right)^{\f{1}{2}}\, \f{1}{\lambda}\ee
This implies that the spectrum of $\hat{A}|_{\phi}$ is bounded above
by the right side of (\ref{Abound}). Therefore the spectrum of
$\hat\rho|_\phi = \hat{A}^2|_{\phi}$ is also bounded by
\be \rho_{\rm sup} = \f{3}{8\pi\gamma^2 G}\,\, \f{1}{\lambda^2} =
\f{\sqrt{3}}{32\pi^2\gamma^3 G^2 \hbar}\,\, \approx 0.41 \rho_{\rm
Pl} \, ,\label{rhosup}\ee
where in the last step we have used the value $\gamma \approx 0.24$
for the Barbero-Immirzi parameter that led to $\rcr \approx 0.41
\rho_{\rm pl}$ in \cite{aps2,apsv,kv}. We wish to emphasize that
this is an absolute bound on the spectrum of $\hat\rho|_{\phi}$ on
the entire physical Hilbert space $\Hp$; there is no restriction
that the states be, e.g., semi-classical. Note also that a factor of
$10$ in the value of $\gamma$ would change $\rsup$ (and $\rcr$) by a
factor of $10^3$. The fact that the value of $\gamma$ obtained from
the entropy calculation yields $\rsup \sim \rho_{\rm Pl}$ points to
an overall coherence of LQG.

Finally, we could also have defined, as a measure of mean density,
the quantity $\tilde\rho = \langle\hat{p}_{(\phi)}\rangle^2/2\langle
\hat{V}|_\phi\rangle^2$. It is easy to verify using the reasoning
that led us to (\ref{rhosup}) that $\tilde\rho$ is also bounded
above by $\rsup$.

Two questions arise immediately: Is this supremum attained? And, how
does it relate to the upper bound $\rho_{\B}$ on density \emph{along
individual dynamical trajectories}? The fact that the value of
$\rsup$ is the same as $\rcr$ suggests that the answers to the two
questions are related. This is indeed the case.

Let $\tilde{F}(k)$ be a smooth function satisfying the following
conditions:
\be \tilde{F}(k) = \f{1}{k}\, e^{-\f{\beta^2(k-k_o)^2}{2}}\,
e^{ikx_o}\,\,\, {\rm if}\,\,\,\, k>\epsilon\, \sqrt{G} >0 \quad {\rm
and}\quad \tilde{F}(k)=0 \,\,\,{\rm if}\,\,\, k\le0 \, ,\ee
where $\epsilon \ll 1 \ll k_o/\sqrt{G}$. This is a semi-classical
initial state peaked at $k=k_o$ and $x=x_o$. Let $F(x)$ be its
Fourier transform. Then $\chi(x,\phi) = (1/\sqrt{2})\, (F(x_+) -
F(x_-))$ is a physical state. One can readily calculate the
expectation values $\pphix$ and $\vex$ in this state. Since $\pphix$
is a constant of motion and $\vex$ attains its minimum, $V_{\B}$, at
the bounce point, as expected $\rho$ attains its maximum also at the
bounce point. It is given by:
\be \rho_{\B} = \rsup\,\, \left[1- O\, \left(\f{G\hbar^2}{\pphi^2+
(\Delta\pphi)^2}\right)\,\,\right] \ee
where we have used the fact that $\Delta\pphi = \hbar\beta$.

Thus, for semi-classical states $\rho_{\B}$ is very close to
$\rsup$. Typical numerical simulations of \cite{aps2} used such
semi-classical states with $\pphix = 5\times 10^3 \hbar$ (in the
classical units $G=c=1$) with the relative dispersion $\Delta
\pphi/\pphix$ of 2.5\%. In this case, the above calculation shows
that $\rsup - \rho_{\B} =  O(10^{-4})$, whence the value $\rcr =
0.41 \rho_{\rm Pl}$ of that numerical simulation is consistent with
the above analytical calculation in the simplified model. Note
incidentally that even this state represents a universe that is very
quantum mechanical. In the $k=1$ case for example, a state with
these parameters represents a universe which grows to a
\emph{maximum} radius of only $\sim 25 \lp$ before undergoing a
classical recollapse. If we use values of $\pphix$ and $\Delta\pphi$
that correspond to universes that grow, say a megaparsec size in the
$k=1$ case \cite{apsv}, $\rsup$ would agree to $\rho_{\B}$ to 1 part
in $10^{230}$. These considerations show that on $\Hp$, $\rho_{\B}$
can come arbitrarily close to $\rsup$.

\subsection{ Fundamental discreteness of sLQC}
\label{s5.2}

We saw in section \ref{s4.2} that the \WDW dynamics does not result
in the limit $\lambda \rightarrow 0$ of sLQC${}_{(\la)}$. A natural
question then is whether sLQC${}_{(\la)}$ admits any limit at all as
the area gap shrinks to zero. We will now show that the answer is in
the negative. The underlying conceptual idea is simple: the
one-parameter family of theories sLQC${}_{(\la)}$ will admit a limit
as $\la \rightarrow 0$ if, given any $\epsilon >0$ there exists a
$\delta>0$ such that the predictions of all sLQC${}_{(\la)}$ with
$\lambda < \delta$ agree within $\epsilon$ for all times $\phi$. We
will argue by contradiction.

Suppose then that the limit exists. Fix an $\epsilon$ and let
$\lambda, \lambda' < \delta$. Fix a normalized state
$\chi_{{}_{(\lambda)}}$ in sLQC${}_{(\lambda)}$. Then by assumption
sLQC${}_{(\lambda')}$ admits a normalized state $\chi_{{}_{(\la')}}$
which is `close' to it. In particular then,
\be\label{near} |\vex_{(\lambda)}- \vex_{(\lambda')}| < \epsilon\, ,
\ee
and, to keep transparency of equations we will assume that
$\pphix_{(\lambda)} = \pphix_{(\lambda')}$.%
\footnote{It is completely straightforward to extend the argument to
allow these expectation values to be different but within $\epsilon$
of each other. Requiring an exact equality is also motivated by the
fact that $\pphi$ is a constant of motion and, even within a single
sLQC${}_{(\lambda)}$ theory, semi-classical states with different
values of $\pphi$ depart from each other unboundedly under
evolution.}
Now, since
\be \vex_{(\la)}\, =\, V_+(\la) e^{\co \phi} + V_-(\la) e^{-\co
\phi}\, , \ee
Eq (\ref{near}) can be satisfied for all $\phi$ if and only if the
$V_\pm (\la) = V_\pm(\la')$. It then follows from (\ref{bounce})
that the densities at the bounce points of the two theories are
equal:
\be \rho_{\B}(\lambda)\, = \,\f{1}{8}\, \f{\pphix^2}{V_+{(\la)}
V_-{(\la)}}\, =\, \f{1}{8}\, \f{\pphix^2}{V_+{(\la')} V_-{(\la')}}
= \rho_{\B}(\lambda') ~.\label{5.10}\ee
Now fix $\la' <\delta$, choose  $\lambda = \la'/N$ for some $N>1$
and let $\chi_{{}_{(\lambda)}}$ be a semi-classical state with
$\rho_{B}(\la) = \rsup(\la) - \t\epsilon$.  As we saw in section
\ref{s5.1}, we can choose $\chi_{{}_{(\lambda)}}$ so that
$\t\epsilon$ is arbitrarily small. But then (\ref{5.10}) and
(\ref{rhosup}) imply
\be \rho_{B}(\lambda')\, = \,\rho_{B}(\lambda)\,=
\, \rsup(\la) - \t\epsilon = N^2\,
\rsup(\la') - \t\epsilon \,\,> \,\,\rsup(\la') \ee
if we choose $N$ to be sufficiently large. But this is impossible
since $\rsup(\la')$ is an absolute upper bound on density in the
physical Hilbert space of the sLQC${}_{(\lambda')}$ theory. This
implies that our initial assumption that the limit exists is
invalid. sLQC does not admit a continuum limit; it is a
fundamentally discrete theory. This establishes the main result of
this section.

It is useful to note that, by replacing $\b_o\lambda$ with
$\la/\lambda'$ in (\ref{flow}), the procedure introduced in section
\ref{s4.2} leads to a rather natural `renormalization flow' relating
states of the 1-parameter family of sLQC${}_{(\lambda)}$ theories.
Under this flow, if the initial $\chi_{{}_{(\la)}}(x)$ is
semi-classical in sLQC${}_{(\la)}$, the state
$\chi_{{}_{(\la')}}(x)$ will also be semi-classical in
sLQC${}_{(\la')}$. The classical trajectories on which the two
states are peaked in the future will be very close to one another.
Each state will also remain peaked on a classical trajectory in the
(pre-bounce) distant past. However, these two classical trajectories
will be very different and diverge away from each other in the past
evolution. Thus, while there is a bounce in each
sLQC${}_{(\lambda)}$, the pre-bounce dynamics is vastly different
for different values of $\lambda$.

The non-convergence of sLQC in the $\la\rightarrow 0$ limit shows
that sLQC is qualitatively different from polymer quantum mechanics
\cite{afw,cvz,cvz2} and lattice gauge theories. The polymer particle
example was introduced as a toy model to probe certain mathematical
and conceptual issues and by itself does not have direct physical
significance. While it has certain similarities with LQC, there are
also important differences. Because there is no positive frequency
requirement in the polymer particle theory, the physical inner
product and the Hamiltonian are local. Arguments that are successful
in that example simply break down because of non-locality of sLQC.
Lattice gauge theories are meant to be controlled approximations to
the continuum theory. As in numerical analysis of PDEs, the
discreteness is introduced just as an intermediate mathematical
simplification. Therefore, absence of a continuum limit would be
tantamount to non-viability of that theory. In LQG on the other hand
the continuum is only an approximation. Not only is there no a
priori reason for the theory to make sense as the area gap is shrunk
to zero, but within the LQG framework it would be hard to make
physical sense of the limit if it existed. Discreteness at the
Planck scale is a fundamental and essential ingredient of the
theory. The continuum emerges on coarse graining; ignoring the fine
structure of quantum geometry because one is not interested in
phenomena at the Planck scale is very different from taking the
naive continuum limit $\lambda \rightarrow 0$ which corresponds to
washing out quantum geometry at all scales.

\section{Discussion}
\label{s6}

Detailed analysis of the physical sector of LQC has revealed
that the older quantum Hamiltonian constraints \cite{mb1,abl} have
serious drawbacks already in the FRW models coupled to at least one
massless scalar field \cite{aps1}. In particular, they lead to
physically unacceptable breakdown of general relativity in
completely `tame' regimes. The `improved' constraint operator of
\cite{aps2} is free of these drawbacks. Within the FRW models, this
improvement is robust in the sense that it extends to situations in
which there is a non-zero cosmological constant and/or non-zero
spatial curvature corresponding to the $k$=1 cosmologies
\cite{aps2,apsv}. Recent analysis has shown that its bounce picture
is also robust with respect to the inclusion of a phenomenologically
viable inflationary potential \cite{aps3}. Thus, the suggestion
(see, e.g., \cite{kks,mb2}) that the bounce would not persist once a
potential is included has turned out to be incorrect.

However, so far predictions from `improved' dynamics were all made
numerically using states which are semi-classical at late times.
Three somewhat different classes of semi-classical states were used
and numerical simulations were carried out in two distinct ways,
using a variety of values of the parameters involved. While these
features had established a certain degree of robustness of results,
in absence of an analytical understanding of dynamics, it was not
obvious whether the results would continue to hold for generic
states. Indeed, it has been suggested that they may not \cite{mb2}
and concerns have been expressed that numerical subtleties may
affect the robustness of the bounce \cite{pl}.

In this paper we obtained an analytically soluble model ---sLQC---
by adapting the theory to the scalar field clock already at the
classical level. This allows one to consider generic states and
analyze physics without recourse to any numerics. The question then
is: Does the quantum bounce persist generically and, if so, does it
continue to be approximately symmetric about the bounce point? Or,
are these features restricted just to semi-classical states? A
natural avenue to explore these questions is through dynamics of the
expectation values of the volume operator. We found that these
undergo a quantum bounce for \emph{all} states and furthermore the
bounce continues to be symmetric. Thus the bounce and its salient
qualitative feature are quite robust in the cosmological model under
consideration. More importantly, we could show that matter density
has an absolute supremum $\rho_{\rm sup}$ on the physical Hilbert
space, given by: $\rho_{\rm sup} = \sqrt{3}/ 32\pi^2 \gamma^3
G^2\hbar \approx 0.41 \rho_{\rm Pl}$. This is precisely the value of
the critical density $\rcr$ \cite{aps2,apsv} at which the bounce
occurs in numerical simulations and in solutions to effective
equations, both of which are, however, based on states that are
semi-classical at late times. Thus the raison d'\^{e}tre and the
observed robustness of the somewhat mysterious $\rcr$ was clarified
analytically.

We could also use the model to analyze the relation between the \WDW
theory and sLQC. It is well known that the key differences between
the two theories can be traced back to the fact that dynamics of LQC
incorporates the quantum nature of geometry through the area gap
$\Delta \, \lp^2$. The question then is: Can one regard the \WDW
theory as the limit of LQC when quantum geometry effects are ignored
by taking the mathematical limit $\Delta \rightarrow 0$? Since it is
analytically soluble, sLQC is well suited to probe this issue. We
used the complete set of Dirac observables, $\hat{p}_{(\phi)},
\hat{V}|_{\phi}$ to compare the two theories and obtained two
results. First, suppose we fix any {semi-infinite} interval $I$ of
internal time and an $\epsilon>0$. Then one \emph{can} approximate
sLQC by the \WDW theory to within $\epsilon$ over the time interval
interval $I$ simply by shrinking by hand the area gap sufficiently.
In this sense the answer to the question is in the affirmative.
However, if one is interested in the \emph{global} time evolution
---i.e., if we let $I$ be the full real line--- then the answer is
in the negative: No matter how much we shrink the area gap, if we
wait long enough the difference in the predictions of the two
theories will become as large as we want. Furthermore, in terms of
global behavior in time, sLQC fails to admit any well-defined limit
as the area gap shrinks to zero. In this sense sLQC is a
fundamentally discrete theory. The use of a non-zero $\Delta$ in LQC
is not just an intermediate step to make the quantum theory
mathematically manageable. Rather, the presence of a non-zero area
gap is a central physical feature of LQC, an imprint of the quantum
geometry of full LQG on this symmetry reduced theory. The dramatic
difference from the more familiar lattice gauge theories is that
whereas QCD can be meaningfully formulated in the continuum, quantum
general relativity cannot; in LQG it is essential to use quantum
geometry at the Planck scale.

We wish to emphasize, however, that the analysis of this paper has
two main limitations. First, we restricted ourselves to the simplest
model, the $k$=0 FRW cosmology coupled to a massless scalar field.
The robustness refers to states within this model (although most
results can be probably generalized by including a cosmological
constant or allowing the $k$=1 closed models). The second and much
more important limitation is that as of now LQC has \emph{not} been
systematically derived from LQG. In the LQC Hamiltonian constraint,
the area gap enters through the definition of the field strength
$F_{ab}^i$ via holonomies of the gravitational connection $A_a^i$.
This strategy is standard in full LQG. However, in LQC one breaks
diffeomorphism invariance through gauge fixing. While this is the
standard practice in all approaches to (classical and) quantum
cosmology, it implies that we cannot directly employ the full
strategy used so far in LQG. The new element ---which may also be
useful in a suitably gauge fixed version of full LQG--- is that the
loop along which the holonomy is defined is shrunk not to zero but
only till it encloses an area $a_o$ of Planck size. Now the most
natural value of $a_o$ is the smallest non-zero area eigenvalue
$\Delta \, \lp^2$ in the class of states relevant to LQC, called the
area gap. However, in this step one parachutes by hand a result from
full LQG into LQC. This is analogous to the procedure used in the
Bohr atom where one puts in the quantization $j = n\hbar$ of angular
momentum by hand. In retrospect, this step is parachuted from the
more complete quantum mechanics. However, a fuller understanding
shows that while angular momentum is indeed quantized, its
eigenvalues are $\sqrt{j(j+1)}\, \hbar$ rather than $n\hbar$. In a
similar vein the quantization of area is parachuted into LQC from
the more complete theory of quantum geometry in LQG. We believe that
this step will be eventually justified through a systematic
derivation of LQC from LQG. However, just as the correct eigenvalues
of angular momentum are more subtle than the `natural' or `obvious'
values $n\hbar$ used in the Bohr atom, the correct value of the area
$a_o$ to which the holonomy loops have to be shrunk may have a value
different from $\Delta \, \lp^2$. For instance, arriving at LQC from
LQG may well require a coarse graining which could lead to a
`dressing' of $\Delta$.

We will conclude by comparing and contrasting our model with another
soluble model that has recently appeared in the literature also for
the $k=0$ cosmology with a massless scalar field \cite{mb2}. There, in
essence one begins with a classical constraint, $c^2p^2 = {\rm
const}\,\, \pphi^2$, takes its positive square root and quantizes by
replacing $c$ by $\sin c$ and $p$ by $-i\hbar \p_c$. Thus one first
simplifies the \emph{classical} constraint and then quantizes,
using rules motivated by LQC.%
\footnote{The replacement $c \rightarrow \sin c$ is motivated by the
older, `$\mu_o$ quantization' of LQC \cite{mb1,abl,aps1} but with
$\mu_o$ set to 1. However, since it is just $c$ rather than $c^2$
that now appears in the square root of the classical constraint, it
cannot be directly related to the field strength $F_{ab}^i$, whence
the replacement $c \rightarrow \sin c$ no longer `descends' from
LQG. See the remark at the end of section \ref{s3}.}
By contrast, in this paper started with the lapse function suited to
harmonic rather than proper time but then followed the same
procedure as in \cite{aps2} to obtain the improved \emph{quantum}
constraint. Therefore, reservations \cite{mb2} that the simplified
model may leave out interesting physics are inapplicable for the
model discussed in this paper. At least for the $k$=0 FRW models
with a massless scalar field, the bounce scenario that emerged from
numerical simulations in \cite{aps2} appears to be robust, i.e., is
not tied to states which are semi-classical at late times.

Finally, there are three other key differences between the two
exactly soluble models. First, the simplified constraint used in
\cite{mb2} corresponds to a difference equation with uniform steps
in $p \sim a^2$ rather than in volume ($\sim a^3$). Therefore, it is
the analog of the older Hamiltonian constraint \cite{mb1,abl} used
in LQC which, as we discussed in sections \ref{s1} and \ref{s3.1},
leads to physically unacceptable predictions. Second,
simplifications of \cite{mb2} removes, by hand, the right or the
left moving sector of the theory. This truncation is motivated by
considerations of mathematical simplicity but does not appear to
have any physical justification. Third, as in LQC, the `evolution
equation' satisfied by physical states in our simplified model
involves the square-root of a positive operator ($\Theta$) and is
thus fundamentally non-local. This is a direct consequence of the
group averaging procedure used to obtain the physical Hilbert space.
By contrast, because the square-root of the constraint is taken
classically in \cite{mb2}, the resulting operator is local (in $c$)
unlike the corresponding operator in full LQC \cite{aps2}. In the
terminology of section \ref{s3}, while the physical states of the
model discussed in this paper have only `positive frequency' modes
but with both right and left moving components, those of \cite{mb2}
contain both positive and negative frequency modes but restricted to
be either left or right moving.

The simplified model introduced in this paper will be used in
\cite{cs} also to analyze properties of fluctuations in volume.
Again it will turn out that indications provided by numerical
simulations involving states which are semi-classical at late times
\cite{aps2} are realized by generic states.\vfill\break

\textbf{Acknowledgments:} We thank Jerzy Lewandowski, Tatjana
Vukasinac, Jos\'e A. Zapata and especially  Tomasz Pawlowski for
discussions. We are grateful to Ian Lawrie for communicating to us
his careful analysis which made us realize that we had mixed two
inequivalent notions of square roots of operators in an intermediate
step of the last version of this paper. Although the final results
remain unaffected, this observation led to intermediate corrections
which have added conceptual clarity. This work was supported in part
by the CONACyT grant U47857-F and NSF grant PHY04-56913, the
Alexander von Humboldt Foundation, the Eberly research funds of Penn
State and the AMC-FUMEC exchange program. The research of PS is
supported by Perimeter Institute for Theoretical Physics.  Research
at Perimeter Institute is supported by the Government of Canada
through Industry Canada and by the Province of Ontario through the
Ministry of Research \& Innovation.

\begin{appendix}

\section{Starting with harmonic versus proper time already in
the classical theory} \label{a1}

In the `improved' dynamics of \cite{aps2}, one began with proper
time in the classical theory (where the lapse is given by $N=1$) and
switched to the scalar field time only after obtaining the quantum
Hamiltonian constraint. In section \ref{s3} by contrast we first
adapted the classical theory to the scalar field time (by using $N=
a^3$) and then proceeded with quantization. The two strategies yield
slightly different factor orderings in the quantum Hamiltonian
constraint. In this Appendix we will discuss the detailed relation
between them. It will turn out that the difference is negligible on
physical grounds.

If one begins with the lapse $N=1$ and then passes to quantum
theory, the gravitational part of the `improved' Hamiltonian is
given by (see Eqs. (2.23)-(2.25) of \cite{aps2}):
\be \hat{C}_{\rm grav}\,\tilde{\Psi}(\nu, \phi) = (\sin\lambda\b)\,
A(\nu)\, (\sin\lambda\b)\, \tilde{\Psi}(\nu,\phi) \ee
with \be A(\nu) = -\f{6\pi \lp^2}{\gamma\lambda^3}\, |\nu|\,\,
\big|\, |\nu+\lambda| -|\nu-\lambda|\, \big|\, , \ee
where  $\lambda>0$ is given by $\lambda^2 = \Delta\,\,\lp^2 \equiv
2\sqrt{3} \pi \gamma \,\lp^2$. The matter part of the Hamiltonian
constraint is given by:
\be \hat{C}_{\rm matt}\,\tilde{\Psi}(\nu, \phi))  = 8\pi G\,
\hat{p}_{(\phi)}^2\, \widehat{V^{-1}}\, \tilde{\Psi}(\nu,\phi)\,
.\ee
Using the Thiemann strategy, \cite{tt,ttbook}, we can calculate the
operator $\widehat{V^{-1}}$ in the $\nu$-representation. As
expected, this operator is diagonal:
\ba \widehat{V^{-1}}\,\tilde{\Psi}(\nu) &=& \f{27}{64}\,
\f{1}{\lambda^3\alpha}\, \big|\, |\nu + \lambda|^{\f{2}{3}} - |\nu -
\lambda|^{\f{2}{3}}\,\big|^3\, \tilde{\Psi}(\nu,\phi) \nonumber\\
&=:& B(\nu) \tilde{\Psi}(\nu,\phi)\, .  \ea
Thus, if one begins with $N=1$ in the classical theory, the
Hamiltonian constraint is given by
\be \label{hc10}\p_\phi^2 \t\Psi(\nu,\phi) = \f{1}{8\pi G \hbar^2}\,
B^{-1}(\nu)\, \sin\lambda\b \, A(\nu)\, \sin\lambda\b \,
\t\Psi(\nu,b)\ee
The key difference from the constraint (\ref{hc4}) used in the main
text is that in (\ref{hc4}) each of the functions $A(\nu)$ and
$(B(\nu))^{-1}$ is replaced by (certain multiples of) $|\nu|$.

\begin{figure}[tbh!]
\includegraphics[scale=0.6]{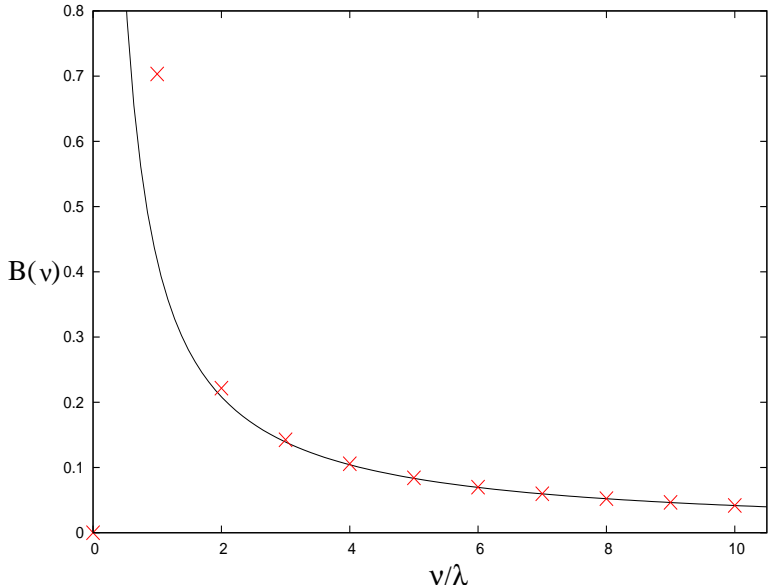}
\caption{Crosses denote values of the LQC function $B(\nu)$
for $\nu/\lambda=n$. The continuous curve represents the sLQC
approximation used in this paper. Physically relevant points are
$\nu/\la =4n$ with $n \ge 1$. The relative `error' is $1.43$\% for
$n=1$, $0.02$\% for $n=2$ and further decreases extremely rapidly
for higher $n$.}\label{fig:1}
\end{figure}

The question then is: Under what conditions does (\ref{hc10}) reduce
to (\ref{hc4})? Note first that in the main text we restricted
ourselves to the lattice $\nu= 4n \lambda = 4n \sqrt{\Delta} \lp$.
On the points of this lattice, we have $A(\nu)= -(12\pi \lp^2/
\gamma\lambda^2)\, |\nu|$. For $B(\nu)$ let us make a weak
assumption:
\begin{itemize}

%
%

\item  \textit{Let us replace $B(\nu)$ with its} {\rm \WDW}
    \textit{value $(2\pi\gamma\lp^2\,|\nu|)^{-1}$.} This amounts
    to assuming $O(|\lambda/\nu|) \ll 1$. Now, because of the
    form of the inner product, the state $|\nu\rangle=0$ on
    which the approximation would have been the worst does not
    feature in the physical Hilbert space of \cite{aps2} nor in
    the physical Hilbert space used in the main text. As
    figure~\ref{fig:1} shows the `error' is only 1.43\% for
    $\nu= 4\lambda$, 0.02\% for $\nu = 8\lambda$ and decreases
    extremely rapidly for higher $\nu$.

\end{itemize}

With this simplification%
\footnote{In \cite{warsaw2} properties of the Hamiltonian constraint
of \cite{aps2} were established using a procedure which began
precisely with our simplified operator. That analysis also provides
a systematic leading order correction to the `simplification' made
here.},
the total constraint $(\hat{C}_{\rm grav} + \hat{C}_{\rm matt})
\t\Psi\, =\,0$ takes the form
\be 
\partial_\phi^2\, \tilde{\Psi}(\nu,\phi) = 3\pi G\, |\nu|\,
\f{\sin\lambda\b}{\lambda}\, |\nu|\, \f{\sin\lambda\b}{\lambda}\,
\tilde{\Psi}(\nu,\phi)\, \ee
which is precisely the constraint (\ref{hc5}) used in the main text.
To summarize, mathematically the Hamiltonian constraint of
\cite{aps2} reduces to the one used in the main text if one replaces
$B(\nu)$ in \cite{aps2} by its \WDW value. The difference between
the two constraint is less than 2\% even for the state concentrated
at $\nu=4\lambda$ and decreases extremely rapidly for states with
support at higher values of $\nu$.

Finally, it is natural to ask if the simplifying assumption on
$B(\nu)$ is violated near the bounce point. If the violation is
significant, then the conclusions drawn from the solvable model
could be qualitatively different from those drawn from the
constraint used in \cite{aps2}. Now, since in the $\nu$
representation \emph{physical} states have support only on points
$\nu = 4n\lambda$ with $|n|>0$, it follows that $V_{\rm min}$ is
necessarily bounded below by $8\pi \gamma \lp^2\lambda$. This value
is in fact attained by the physical state $\tilde\chi(\nu,\phi) =
(\exp i\sqrt{\Theta}\phi)\,\, \delta_{|n|,1}$ in the
$\nu$-representation, or, $\chi (x,\phi) = (\exp
i\sqrt{\Theta}\phi)\,\,\sin (4\,\arctan \co\,x)$ in the $x$
representation. In this state, the expectation value $\langle \,
\hat{p}_{(\phi)}\rangle$ of the constant of motion
$\hat{p}_{(\phi)}$ is $\sim \hbar$ (in the classical units
$G$=$c$=1). Therefore this state belongs to the extreme quantum
regime. For example, in the case of $k$=1, closed universes,
classical Einstein's equations imply that a universe with such a
value of $\pphi$ can expand out to a \emph{maximum} radius only of
$\sim 0.3\lp$ before recollapsing to a big-crunch. Even for such an
extreme quantum state, the treatment of the main text is close to
that in \cite{aps2}: as figure 1 shows, for $B(\nu)$ the relative
error is $\sim 1.4\% $. In any state which can be thought of as
representing a classical universe at late times even in a very weak
sense, the difference between the two factor orderings would be
further suppressed by an enormous factor.

\end{appendix}


\begin{thebibliography}{99}

\bibitem{apslett} A.~Ashtekar, T.~Pawlowski and P.~Singh, {Quantum
    nature of the big bang}, Phys. Rev. Lett. \textbf{96}, 141301
    (2006),
 \texttt{arXiv:gr-qc/0602086}.

\bibitem{aps1} A.~Ashtekar, T.~Pawlowski and P.~Singh, {Quantum
    nature of the big bang: An analytical and numerical
    investigation I}, Phys. Rev. {\bf D73}, 124038,
    \texttt{arXiv:gr-qc/0604013}.

\bibitem{aps2} A.~Ashtekar, T.~Pawlowski and P.~Singh, {Quantum
    nature of the big bang: Improved dynamics}, Phys. Rev. {\bf
    D74}, 084003, \texttt{arXiv:gr-qc/0607039}.

\bibitem{apsv} A.~Ashtekar, T.~Pawlowski, P.~Singh and
    K.~Vandersloot, {Loop quantum cosmology of $k$=1 FRW models},
    Phys. Rev. \textbf{D75}, 0240035 (2007),
    \texttt{arXiv:gr-qc/0612104}.

\bibitem{awe2} A.~Ashtekar and E.~Wilson-Ewing, {Loop quantum
    cosmology of Bianchi type I models}, Phys. Rev. \textbf{D79}083535 (2009).

\bibitem{warsaw} L.~Szulc, W.~Kaminski, J.~Lewandowski, Closed FRW
    model in loop quantum cosmology, \texttt{arXiv:gr-qc/0612101}.

\bibitem{aarev} A.~Ashtekar, An Introduction to Loop Quantum Gravity
    through Cosmology, Nuovo Cimento \textbf{112B}, 1-20 (2007),
    \texttt{arXiv:gr-qc/0702030}

\bibitem{bt} J.~Brunnemann and T.~Thiemann, {On(cosmological)
    singularity avoidance in loop quantum gravity},
    Class.\ Quant.\ Grav.\  {\bf 23}, 1395 (2006)
    \texttt{arXiv:gr-qc/0505032}

\bibitem{gu} D.~Green and W.~Unruh, Difficulties with recollapsing
    models in closed isotropic loop quantum cosmology, Phys. Rev.
    \textbf{D70}, 103502 (2004), \texttt{arXiv:gr-qc/0408074}.

\bibitem{mb1} M.~Bojowald, {Absence of singularity in loop quantum
    cosmology}, Phys. Rev. Lett. \textbf{86}, 5227-5230
(2001), \texttt{arXiv:gr-qc/0102069};\\
{Isotropic loop quantum cosmology}, Class. Quantum. Grav.
\textbf{19}, 2717-2741 (2002), \texttt{arXiv:gr-qc/0202077}.

\bibitem{abl} A.~Ashtekar, M.~Bojowald, J.~Lewandowski,
    {Mathematical structure of loop quantum cosmology}, Adv. Theo.
    Math. Phys. \textbf{7}, 233-268 (2003), \texttt{gr-qc/0304074}.

\bibitem{bck} M.~Bojowald, D.~Cartin and G.~Khanna,
 Lattice refining loop quantum cosmology, anisotropic models and
  stability,
  Phys.\ Rev.\  D {\bf 76}, 064018 (2007)
  \texttt{arXiv:0704.1137 [gr-qc]}

\bibitem{kks} A.~Kamenshchik, C.~Keifer and B.~Sandhofer, Quantum
    cosmology with big brake singularity, Phys. Rev. \textbf{D76},
    064032 (2007), \texttt{arXiv:0705.1688 [gr-qc]}.

\bibitem{kv} K.~Vandersloot,
  Loop quantum cosmology and the $k = -1$ FRW model,
  Phys.\ Rev.\  D \textbf{75}, 023523 (2007)
  \texttt{arXiv:gr-qc/0612070}.

\bibitem{mb2} M.~Bojowald, Large scale effective theory for
    cosmological bounces, Phys. Rev. \textbf{D74}, 081301 (2007),
    \texttt{arXiv:gr-qc/0608100};\\
    What happened before the big-bang? Nature Physics \textbf{3},
    523-525 (2007).

\bibitem{afw} A.~Ashtekar, S.~Fairhurst and J.~Willis, {Quantum
    gravity, shadow states, and quantum mechanics}, Class.\ Quantum\
    Grav. \textbf{20}, 1031-1062 (2003),
    \texttt{arXiv:gr-qc/0207106}.

\bibitem{cvz} A.~Corichi, T.~Vukasinac and J.~A.~Zapata,
  Hamiltonian and physical Hilbert space in polymer quantum mechanics,
  Class.\ Quant.\ Grav.\  \textbf{24}, 1495 (2007),
  \texttt{arXiv:gr-qc/0610072}

\bibitem{cvz2}
A.~Corichi, T.~Vukasinac and J.A.~Zapata, Polymer
    quantum mechanics and its continuum limit, Phys. Rev.
    \textbf{D76}, 044016 (2007), \texttt{arXiv:0704.0007}

\bibitem{cs}A.~Corichi and P.~Singh, Quantum bounce and cosmic recall,
Phys. Rev. Lett. {\bf 100}, 209002 (2008),
\texttt{arXiv:0710.4543 [gr-qc]}.

\bibitem{hawking} S.W.~Hawking, Quantum cosmology, in
    \textit{Relativity Groups and Topology}, edited by B.S. DeWitt
    and R. Stora (Elsevier, Amsterdam 1983).

\bibitem{as} A.~Ashtekar and J.~Samuel, Bianchi Cosmologies: The
role of spatial topology, Class.\ Quantum\ Grav. \textbf{8},
2191-2215 (1991).

\bibitem{dm} D.~Marolf, Refined algebraic quantization: Systems with
    a single constraint \texttt{arXive:gr-qc/9508015}. 


\bibitem{abc} A.~Ashtekar, L.~Bombelli and A.~Corichi, Semiclassical
    states for constrained systems, Phys. Rev. \textbf{D72}, 025008
    (2005), \texttt{gr-qc/0504052}.

\bibitem{almmt} A.~Ashtekar, J.~Lewandowski, D.~Marolf, J.~Mour\~ao
    and T.~Thiemann, Quantization of diffeomorphism invariant theories
    of connections with local degrees of freedom {Jour. Math. Phys.}
    \textbf{36} 6456--6493 (1995).

\bibitem{tt} T. Thiemann,
Quantum spin dynamics (QSD), {Class. Quant. Grav.} \textbf{15}
839--873 (1998), \texttt{gr-qc/9606089}, \\ QSD V : Quantum gravity
as the natural regulator of matter quantum field theories, {Class.
Quant. Grav.} \textbf{15}, 1281--1314 (1998),
\texttt{gr-qc/9705019}.

\bibitem{ttbook} T.~Thiemann, {\em Modern Canonical
    Quantum General Relativity} (Cambridge University Press, Cambridge, 2007).

\bibitem{ai} A.~Ashtekar and C.J.~Isham, Representation of
    the holonomy algebras of gravity and non-Abelian gauge theories
    \textit{Class. Quant. Grav.} \textbf{9}  1433--1467 (1992).

\bibitem{lost} J.~Lewandowski, A.~Okolow, H.~Sahlmann and
    T.~Thiemann, {Uniqueness of diffeomorphism invariant states on
    holonomy flux algebras},  Comm. Math. Phys. \textbf{267},
    703-733 (2006) \texttt{arXiv:gr-qc/0504147};\\
    C.~Fleischhack, {Representations of the Weyl algebra in quantum
    geometry}, \texttt{arXiv:math-ph/0407006}.

\bibitem{warsaw2} W.~Kaminski and J.~Lewandowski,
  The flat FRW model in LQC: the self-adjointness,
  \texttt{arXiv:0709.3120 [gr-qc]}

\bibitem{other} V.~Husain and O.~Winkler,
 Semiclassical states for quantum cosmology,
  Phys.\ Rev.\  D {\bf 75}, 024014 (2007)
  \texttt{arXiv:gr-qc/0607097}


\bibitem{pl} P.~Laguna,
  The shallow waters of the big-bang,
  Phys.\ Rev.\  D \textbf{75}, 024033 (2007)
  \texttt{arXiv:gr-qc/0608117}


\bibitem{aps3} A.~Ashtekar, T.~Pawlowski and P.~Singh, in
    preparation.

\end{thebibliography}
\end{document}